\documentclass[12pt]{iopart}
\usepackage{graphics}
\usepackage{iopams}  
\begin{document}

\topical{Interplay between electron-phonon and Coulomb interactions 
in cuprates}

\author{O. Gunnarsson\dag\ and O. R\"osch\dag\ \ddag\ }

\address{\dag\ Max-Planck-Institut f\"ur Festk\"orperforschung,
D-70506 Stuttgart, Germany}

\address{\ddag\ Institut f\"ur Theoretische Physik, Universit\"at zu K\"oln,
D-50937 K\"oln, Germany}

\begin{abstract}
Evidence for strong electron-phonon coupling in high-$T_c$ cuprates is  
reviewed, with emphasis on the electron and phonon spectral 
functions.  Effects due to the interplay between the Coulomb
and electron-phonon interactions are studied. For weakly doped 
cuprates, the phonon self-energy is strongly reduced due to correlation
effects, while there is no corresponding strong reduction for the electron 
self-energy. Polaron formation is studied, focusing on effects of Coulomb 
interaction and antiferromagnetic correlations. It is argued that 
experimental indications of polaron formation in undoped cuprates are 
due to a strong electron-phonon interaction for these systems.

\end{abstract}



\maketitle

\section{Introduction}\label{sec:intro}

High-$T_\mathrm{c}$ cuprates show a large number of interesting features, 
apart from the exceptionally large superconducting transition temperature 
$T_\mathrm{c}$. They exhibit antiferromagnetic \cite{AF}, pseudogap 
\cite{Timusk}, marginal Fermi liquid \cite{Marginal} and ordinary Fermi 
liquid phases in addition to the superconducting phase. After the 
high-$T_\mathrm{c}$ cuprates had been discovered \cite{Bednorz}, there 
was initially much interest in the electron-phonon interaction 
(EPI). It was, however, soon concluded that the EPI is too 
weak to explain superconductivity alone, in particular $d$-wave 
superconductivity,  and the interest focused on purely electronic 
models of these compounds. More recently, there has been substantial 
experimental evidence that the EPI plays an appreciable role for 
a number of properties. Certain phonons show a large softening and 
broadening under doping \cite{Pint1,Pint2}, suggesting a strong 
interaction with doped holes. This is, for instance, seen for the 
so-called half-breathing copper-oxygen bond stretching phonon,  
apical oxygen phonons and the oxygen B$_{1g}$ buckling phonon. 
Photoemission spectroscopy (PES) experiments show the formation of 
small polarons for the undoped cuprates \cite{Khyle2004}, and a kink 
in the nodal ${\bf k}$-direction also suggests strong EPI 
\cite{Lanzara2001}. While there is only a weak isotope effect on 
$T_\mathrm{c}$ for optimally doped samples, a strong isotope effect has 
been seen away from optimum doping \cite{Franck1}. Recent STM
work suggests that a phonon mode plays a role in superconductivity
\cite{STMDavis}, although other interpretations are possible \cite{Sigrist}. 
In particular, an isotope effect has been observed \cite{STMDavis}.
While the phonon contribution to superconductivity remains unclear, 
it seems clear that phonons can be important for other properties.  

The EPI has been studied very extensively in the local density approximation 
(LDA) \cite{KS} of the density functional formalism \cite{HK}, which 
is particularly appropriate for systems where correlation effects are 
not very strong. This approach has been shown to be very successful for 
conventional superconductors \cite{Savrasov1,Savrasov2,Gross1,Gross2}. 
For cuprates \cite{Pickett} a rather weak EPI was found, which 
alone would not be sufficient to explain the superconductivity 
\cite{Savrasov3}. However, the calculated width \cite{Bohnen} of the 
half-breathing phonon is an order of magnitude smaller than the 
reported experimental value \cite{Pint4}, raising some questions about
the accuracy of the LDA in this context \cite{threeband}. This is one 
of the reasons that the interest has recently focused on whether the interplay 
between the Coulomb interaction and the EPI can explain experimental 
signs of a strong EPI. 

Due to the important effects of the Coulomb interaction in 
these systems, models such as the Hubbard and $t$-$J$ models 
are often used. In these models important phonons couple to 
charge fluctuations. Since charge fluctuations are strongly
suppressed in the cuprates by the Coulomb interaction, an
important issue is if this could mean that the EPI is actually
suppressed. We discuss this issue extensively below. 

In the so-called sudden approximation, angular resolved photoemission
spectroscopy (ARPES) can be directly related to the one-electron Green's
function. If superconductivity is due to bosons coupling to electrons
and forming electron pairs, this coupling should show up in the 
one-electron Green's function. Due to the high energy- and 
${\bf k}$-resolution that can now be obtained in ARPES, a lot
of interest has focused on ARPES recently, and we address
these issues below.

ARPES experiments strongly indicate that small polarons are formed 
for undoped cuprates and there are signs of strong phonon side bands 
\cite{Khyle2004}. This indicates that there is a strong EPI for these 
systems. For weakly underdoped or optimally doped cuprates, ARPES 
experiments show quasiparticles, suggesting that there are no small 
polarons formed in these cases. However, there is still substantial
spectral weight in the energy range where phonon side bands would be
expected, suggesting that the EPI is still substantial. There has 
been extensive work on polarons and bipolarons in metals, treating 
both electronic properties in general and superconductivity \cite{Alexandrov}. 
Since experiment suggest that small polarons are not formed at dopings
relevant for superconductivity,  we here focus on polaron formation for 
insulating systems. 

Due to the great interest in cuprates, there have been many reviews covering
different aspects of these systems
\cite{AF,Timusk,Pint1,Pint2,Lanzara2001,Franck1,Pickett,Dagotto,Egami1996,Tohyama,Kivelson,Norman2003,Fink2005,Norman2005,Lee,Zhou2006,Yoshida2006,Egami2006}.

\section{Models}\label{sec:m}

\subsection{Coulomb interaction and hopping}

The Coulomb interaction plays an important role in the cuprates.
A frequently used model for describing this is the three-band 
model \cite{Emery}, which includes a Cu $x^2-y^2$ $3d$
orbital and two O orbitals in a CuO$_2$ plane. The model
includes the Cu-O hopping integrals and the Coulomb interaction 
between two electrons on the Cu orbital. 
\begin{eqnarray}\label{eq:m0}
H_{\rm three-band}&=&\varepsilon_\mathrm{d}\sum_{i\sigma}c^{\dagger}_{i\sigma}
c^{\phantom \dagger}_{i\sigma}+
\varepsilon_\mathrm{O}\sum_{i{\bf \delta}\sigma}a^{\dagger}_{i{\bf \delta}\sigma}
a^{\phantom \dagger}_{i{\bf \delta}\sigma} \\
&+&t_\mathrm{pd}\sum_{i{\bf \delta}\sigma}P_{{\bf \delta}}(c^{\dagger}_{i\sigma}
a^{\phantom\dagger}_{i{\bf \delta}\sigma}+\mathrm{H.c.})+
U\sum_i n_{i\uparrow}n_{i\downarrow}, \nonumber
\end{eqnarray}
where $\varepsilon_\mathrm{d}$ and $\varepsilon_\mathrm{O}$ are the energies of the
Cu and O atoms, respectively. ${\bf \delta}$ describes the O atom
positions in the unit cell and runs over $(a/2,0)$ and $(0,a/2)$ 
in the second term and over $(\pm a/2,0)$ and $(0,\pm a/2)$
in the third term, where $a$ is the lattice parameter. $P_{\bf -\delta}=
-P_{\bf \delta}$, $P_{\bf \delta}=1$ for $\delta=(a/2,0)$ and 
$P_{\bf \delta}=-1$ for $\delta=(0,a/2)$. $c_{i\sigma}^{\dagger}$ creates 
a Cu electron in cell $i$ with spin $\sigma$, $a_{i{\bf \delta}\sigma}^{\dagger}$
creates an O electron and $n_{\sigma}=c^{\dagger}_{i\sigma}
c^{\phantom \dagger}_{i\sigma}$. $U$ is the Coulomb interaction and 
$t_\mathrm{pd}$ is a hopping integral.

From this model the $t$-$J$ model can be derived \cite{Zhang}, where 
each site corresponds to a Cu atom in the CuO$_2$ plane. In the undoped 
system, corresponding to all Cu atoms being in $d^9$ configurations, 
each site is occupied by one hole. In a hole doped system, the holes 
go primarily onto the O sites. Such an O hole forms a Zhang-Rice singlet 
with a Cu hole \cite{Zhang}. A Zhang-Rice singlet is described by an 
empty site in the $t$-$J$ model. The corresponding Hamiltonian is 
\begin{equation}\label{eq:m1}
H_{t\textrm{-}J}=-t\sum_{\langle ij\rangle \sigma}(\tilde c_{i\sigma}
^{\dagger} \tilde c_{j\sigma}^{\phantom \dagger} +\mathrm{H.c.})+J\sum_{\langle 
ij\rangle} ({\bf S}_i\cdot{\bf S}_j-{1\over 4}n_in_j),
\end{equation}
where $\langle ij\rangle$ 
refers to a sum over nearest neighbor pairs, and $\tilde c_{i\sigma}
^{\dagger}$ creates a spin $\sigma$ hole on site $i$ if this site 
was previously empty. ${\bf S}_i$ is the spin and $n_i=\sum_{\sigma}
\tilde c_{i\sigma}^{\dagger}\tilde c_{i\sigma}$ is the number of holes 
on site $i$.  

Alternatively, the one-band Hubbard model is often used
\begin{equation}\label{eq:m2}
H_{\rm Hub}=-t\sum_{\langle ij\rangle \sigma}(c_{i\sigma}^{\dagger} 
c_{j\sigma}^{\phantom \dagger}+\mathrm{H.c.}) +U\sum_i n_{i\uparrow}n_{i\downarrow}.
\end{equation}
The $t$-$J$ model can also be derived from the Hubbard model in the
large $U$ limit if certain terms are neglected \cite{Auerbach}.

\subsection{Electron-phonon interaction}
We introduce the Hamiltonian for a set of phonons
\begin{equation}\label{eq:m2a}
H_\mathrm{ph}=\sum_{\bf q}\hbar \omega_{\bf q}b^\dagger_{\bf q}
b^{\phantom\dagger}_{\bf q},
\end{equation}
where $b^{\phantom\dagger}_{\bf q}$ annihilates a phonon with the frequency
$\omega_{\bf q}$ and the wave vector ${\bf q}$.  Generally we write the 
coupling to the phonons as  
\begin{equation}\label{eq:m3}
H_{\rm el-ph}={1\over \sqrt{N}}\sum_{{\bf k}{\bf q}}g({\bf k},{\bf q})
c^{\dagger}_{{\bf k+q}\sigma}c^{\phantom \dagger}_{{\bf k}\sigma}
(b_{\bf q}^{\phantom \dagger}+b_{-{\bf q}}^{\dagger}),
\end{equation}
where $N$ is the number of cells.

\subsubsection{Holstein phonons}

Often, the electron-phonon interaction is treated in a Holstein
model, where there is an on-site coupling to one local Einstein phonon
per site. This corresponds to a ${\bf k}$ and ${\bf q}$ independent
coupling
\begin{equation}\label{eq:m4}
g_{\rm Hol}({\bf k},{\bf q})=g_0,
\end{equation}
where $g_0$ is the coupling constant, and a ${\bf q}$ independent
phonon frequency $\omega_{{\bf q},\mathrm{Hol}}=\omega_\mathrm{ph}$. 

\begin{figure}[bt]
\hskip2.0cm
\begin{minipage}{0.3\linewidth}
{\rotatebox{0}{\resizebox{!}{3.0cm}{\includegraphics {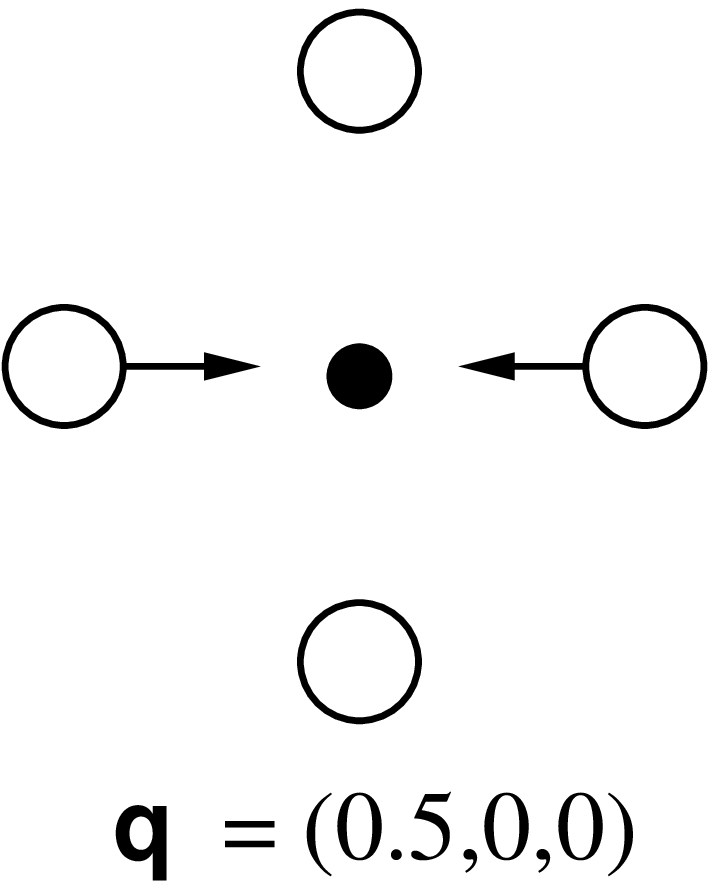}}}}

\vskip0.5cm
{\rotatebox{0}{\resizebox{!}{3.0cm}{\includegraphics {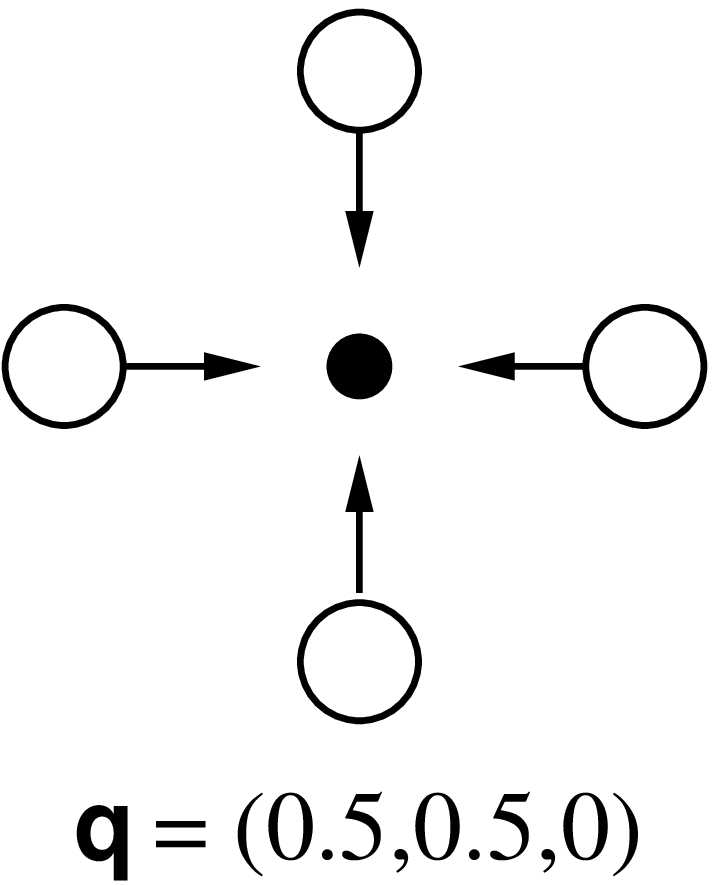}}}}
\end{minipage}
\begin{minipage}{0.3\linewidth}
{\rotatebox{0}{\resizebox{!}{2.5cm}{\includegraphics {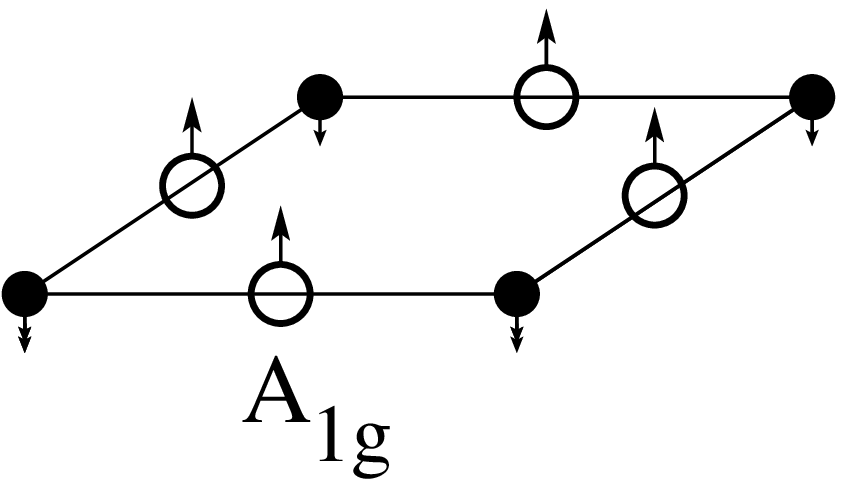}}}}

\vskip0.5cm
{\rotatebox{0}{\resizebox{!}{2.5cm}{\includegraphics {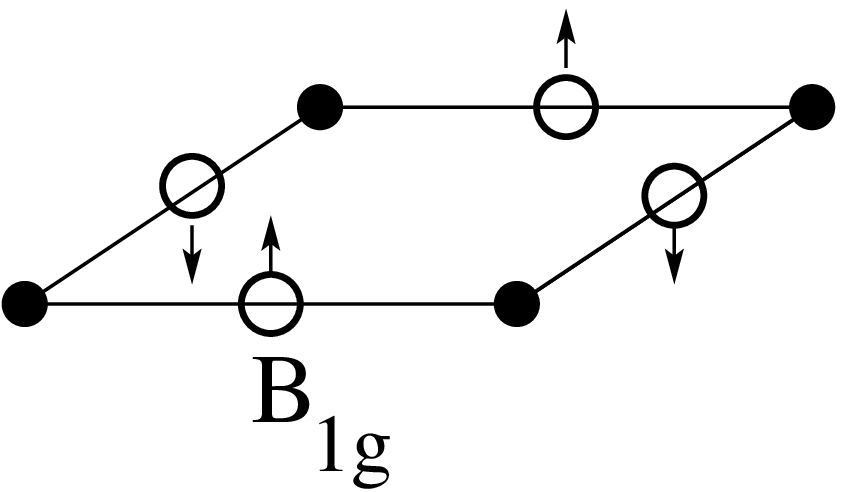}}}}
\end{minipage}
\begin{minipage}{0.2\linewidth}
{\rotatebox{0}{\resizebox{!}{7.0cm}{\includegraphics {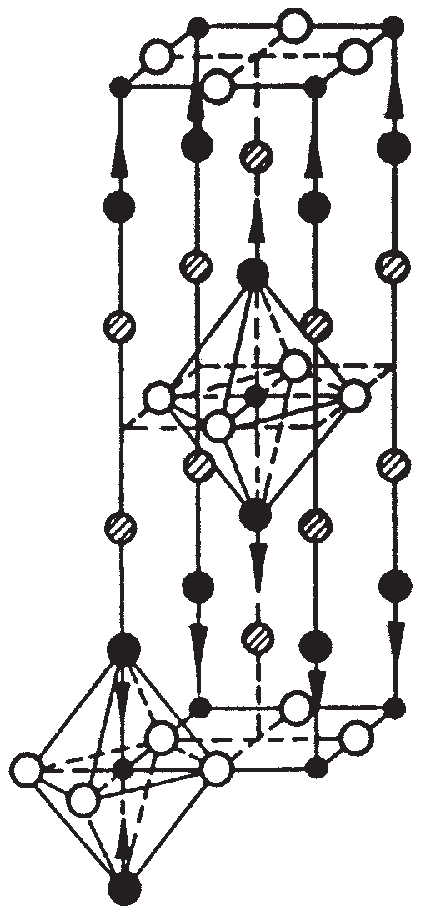}}}}
\end{minipage}
\caption[]{\label{fig:ep2}Half-breathing (upper left), breathing 
(lower left), A$_{1g}$ (upper middle), B$_{1g}$ (lower middle) and apical
oxygen O$_Z^Z$ (right) phonon modes. In the left and middle figures, the 
small filled circles show Cu atoms and the large circles O atoms 
in the CuO$_2$ plane. In the right figure \cite{Pint1}, the large 
black spheres show apical O atoms.
}
\end{figure}

\subsubsection{Breathing phonons}

The breathing (oxygen bond-stretching) phonons have attracted much 
interest due to the observation 
of an anomalous softening and broadening of these phonons when the system 
is doped \cite{Pint1}. That these phonons may have a strong coupling can 
be understood by noticing that the formation of the Zhang-Rice singlet in 
the $t$-$J$ model involves a large energy of the order of several eV. For 
a system without phonons and a fixed number of doped holes, this energy 
only enters as an uninteresting constant. If phonons are added, however, 
the singlet energy can be modulated by the phonons. This is the case for 
the breathing phonons, where the O atoms in the CuO$_2$ plane move
in the direction of the Cu atoms, thereby changing the bond lengths
(see Fig.~\ref{fig:ep2}). This directly modulates the Cu-O hopping 
integrals $t_{\rm pd}$ (in a three-band model) determining the Zhang-Rice singlet 
energy and leads to a substantial coupling. This has been discussed by 
several groups \cite{Becker,Horsch1,Horsch2,Horsch3,Horsch4,Nagaosabreathing,Oliverbreathing}.
A general formula for this coupling was given in Ref. 
\cite{Oliverbreathing}, considering both the modulation of the Cu-O
hopping integrals and shifts of the levels due to Coulomb interactions. 
It was found that the main coupling is an on-site coupling due 
to the modulation of the Cu-O hopping integrals. One reason for this result
is that the hopping integrals in the $t$-$J$ model, obtained after the 
O levels have been projected out, are about an order of magnitude smaller
than the on-site singlet energy. This strongly favors the on-site 
electron-phonon interaction over the coupling to the $t$-$J$ hopping
integrals \cite{onsite}. Below we therefore neglect the off-site 
interaction. If we furthermore assume that the vibration of the Cu atom 
can be neglected due to its larger mass, the oxygen phonon eigenvectors 
can be approximated as $\epsilon_{\alpha}^{\alpha}={\rm sin}(aq_{\alpha}/2)
\sqrt{{\rm sin}^2(aq_x/2) +{\rm sin}^2(aq_y/2)}$, where $\alpha=x$ or $y$, 
describing the motions in the directions of the nearest neighbor Cu atoms. 
Then the coupling becomes
\begin{equation}\label{eq:m4a}
g_{\rm Br}({\bf k},{\bf q})\sim\sqrt{{\rm sin}^2(aq_{\alpha}/2)+
{\rm sin}^2(aq_y/2)},
\end{equation}
i.e., the coupling depends on ${\bf q}$ but not on ${\bf k}$ in this
approximation \cite{Bulut}.

\subsubsection{A$_{1g}$ and B$_{1g}$ phonons}

There has been substantial interest in a B$_{1g}$ phonon at about
42 meV \cite{Cardona2,Pyka,Reznik}, involving a motion
perpendicular to the CuO$_2$ plane of the O atoms in this plane. 
These atoms can move out of phase, leading to a B$_{1g}$ phonon,
or in phase, leading to an A$_{1g}$ phonon (see Fig.~\ref{fig:ep2}). 
Devereaux and coworkers \cite{Devereaux95,Devereaux99,Devereaux2004} 
studied these phonons in the three-band model. They considered the case 
when there is an electrical field, $E_z$, perpendicular to the CuO$_2$
plane, due to different valencies of the ions on the two sides of
a plane. Since the O atoms move perpendicular to the plane for the
A$_{1g}$ and B$_{1g}$ phonons, these phonons couple to such a field. 
For the case of only nearest neighbor hopping, this leads to the
coupling \cite{Devereaux2004} 
\begin{equation}\label{eq:m5}
g_{A_{1g},B_{1g}}({\bf k},{\bf q})\sim E_z\lbrack B_x
\phi^{\ast}_x({\bf k+q})
\phi^{\phantom\ast}_x({\bf k})+B_y\phi^{\ast}_y({\bf k+q})
\phi^{\phantom\ast}_y({\bf k})\rbrack,
\end{equation}
where $\phi_{x\atop y}=\mp i t_{x\atop y}({\bf k})/\sqrt{E^2({\bf k})+
\Omega^2({\bf k})}$, $E({\bf k})=|\varepsilon_\mathrm{d}-\varepsilon_\mathrm{O}|/2+
\sqrt{(\varepsilon_\mathrm{d}-\varepsilon_\mathrm{O})^2/4+\Omega^2({\bf k})}$,
$\Omega^2({\bf k})=t_x^2({\bf k})+t_y^2({\bf k})$ and $t_{\alpha}({\bf k})=
2t{\rm sin}(ak_{\alpha}/2)$. Using a spring model, Devereaux {\it et al.}
\cite{Devereaux99b} obtained the eigenvectors $\lbrack B_x,B_y\rbrack =
\lbrack {\rm cos}(aq_y/2), -{\rm cos}(aq_x/2) \rbrack/M({\bf q})$
for the B$_{1g}$ mode, where $M({\bf q})=\sqrt{{\rm cos}^2(aq_y/2)+
{\rm cos}^2(aq_x/2)}$. Putting the Cu mass equal to infinity, their 
model gives $\lbrack B_x,B_y \rbrack =\lbrack{\rm cos}(aq_x/2), 
{\rm cos}(aq_y/2)\rbrack/M({\bf q})$ for the A$_{1g}$ mode. 

In contrast to the breathing phonons, these modes have a strong ${\bf k}$
dependence. For ${\bf q}=0$, the coupling due to the B$_{1g}$ phonon is 
entirely off-site, while the A$_{1g}$ phonon has a substantial on-site 
coupling. 

A different approach was taken by Jepsen {\it et al.} \cite{Jepsen98}.
They studied a six-band model of the LDA band structure and focused
on the coupling due to the modulation of hopping integrals. This
coupling is zero for a completely flat CuO$_2$ plane but becomes finite
for a dimpled plane. They obtained a coupling
\begin{equation}\label{eq:m6}
g_{B_{1g}}({\bf k},{\bf q})\sim {\rm cos}{ a(k_x+q_x)\over 2} 
{\rm cos}{ ak_x\over 2}
              -{\rm cos}{a(k_y+q_y)\over2} {\rm cos}{ ak_y\over 2}.
\end{equation}
This coupling tends to emphasize small values of $|{\bf k}|$ and $|{\bf k+q}|$
more then Eq.~(\ref{eq:m5}), due to the appearance of cos-functions rather 
than sin-functions in $t_{\alpha}({\bf k})$ entering in Eq.~(\ref{eq:m5}).

\subsubsection{Apical phonons}

Neutron scattering experiments show that several apical oxygen
phonons (see Fig.~\ref{fig:ep2}) broaden and soften when a 
cuprate is doped \cite{Pint1}. This coupling has been calculated 
for La$_2$CuO$_4$ using a shell model \cite{Oliver2005}. Due to 
the ionicity of the O atoms, the electrostatic part of the coupling 
is expected to be particularly strong. This is in particular true for 
the undoped system, which is an insulator, leading to a poor screening
of the electrostatic interaction. It is therefore important not to 
perform this calculation using the LDA, since LDA gives a metallic
system and too efficient screening. Due to the electrostatic nature 
of the coupling, and due to small ${\bf q}$ vectors playing an 
important role, the coupling is expected to be dominated by the 
on-site part of the coupling, i.e.,
\begin{equation}\label{eq:m7}
g_{\rm Apical}({\bf k},{\bf q})\sim g_{\rm A}({\bf q}).                    
\end{equation}
Calculations showed that this coupling is indeed rather strong
\cite{Oliver2005}.

\section{Weak coupling and noninteracting electrons}\label{sec:n}
\subsection{Electron self-energy}\label{sec·na}

The electron-phonon interaction is often studied assuming that the
electrons are noninteracting. This is a quite unrealistic assumption 
for the cuprates, where the electron-electron interaction is crucial. 
Below, we nevertheless describe some of the results \cite{Mahan,Engelsberg,Scalapino1969} 
for this case, since they provide a basis for discussing similarities 
and deviations for strongly correlated systems. The electrons are 
described by the Hamiltonian
\begin{equation}\label{eq:n1}
H_\mathrm{non}=\sum_{{\bf k} \sigma}\varepsilon_{\bf k}c^{\dagger}_{{\bf k}\sigma}
c^{\phantom\dagger}_{{\bf k}\sigma},
\end{equation}
where $\varepsilon_{\bf k}$ is the energy for the wave vector ${\bf k}$ 
and $\sigma$ is a spin index. The electrons are assumed to couple to the phonons 
via the Holstein model (\ref{eq:m4}). We calculate the retarded electron 
self-energy to lowest order in the coupling $g$. For $T=0$, it  is 
given by \cite{Mahan}
\begin{equation}\label{eq:n2}
\Sigma({\bf k},\omega)={1\over N}g^2\sum_{\bf q}\lbrack 
{f(\varepsilon_{\bf k+q}) \over \omega + \omega_\mathrm{ph} -
\varepsilon_{\bf k+q} +i\delta} +{1-f(\varepsilon_{\bf k+q}) 
\over \omega - \omega_\mathrm{ph} -\varepsilon_{\bf k+q}+i\delta}\rbrack,
\end{equation} 
where $N$ is the number of sites, $f(\varepsilon)$ is the Fermi 
function and $\delta$ is a positive infinitesimal (later small) quantity. We 
assume that $N(\varepsilon)=1/B$ is constant, where $N(\varepsilon)$ is the 
density of states (DOS) per spin and $B$ is the band width. The band is 
assumed to be half-filled and to extend from -$B/2$ to $B/2$. Then the 
sum in Eq.~(\ref{eq:n2}) can be performed exactly, giving
\begin{equation}\label{eq:n3}
\Sigma({\bf k},\omega)={1\over 2}\lambda \omega_\mathrm{ph}\lbrack {\rm ln}
{\omega+\omega_\mathrm{ph}+B/2+i\delta \over \omega+\omega_\mathrm{ph}+i\delta}+
{\rm ln} {\omega-\omega_\mathrm{ph}+i\delta \over \omega-\omega_\mathrm{ph}-B/2+i\delta}\rbrack,
\end{equation}
where 
\begin{equation}\label{eq:n3a}
\lambda={2g^2\over \omega_{\rm ph}}N(0). 
\end{equation}
$\Sigma$ is ${\bf k}$-independent
in this approximation. Assuming that $\omega_\mathrm{ph}\ll B/2$, we obtain
\begin{equation}\label{eq:n4}
\Sigma({\bf k},\omega)=\cases {-\lambda \omega, & if $|\omega|\ll \omega_\mathrm{ph}$;\cr
0,& if $\omega_\mathrm{ph} \ll |\omega| \ll B/2$.\cr}
\end{equation} 
\begin{figure}[bt]
{\rotatebox{-90}{\resizebox{!}{9.0cm}{\includegraphics {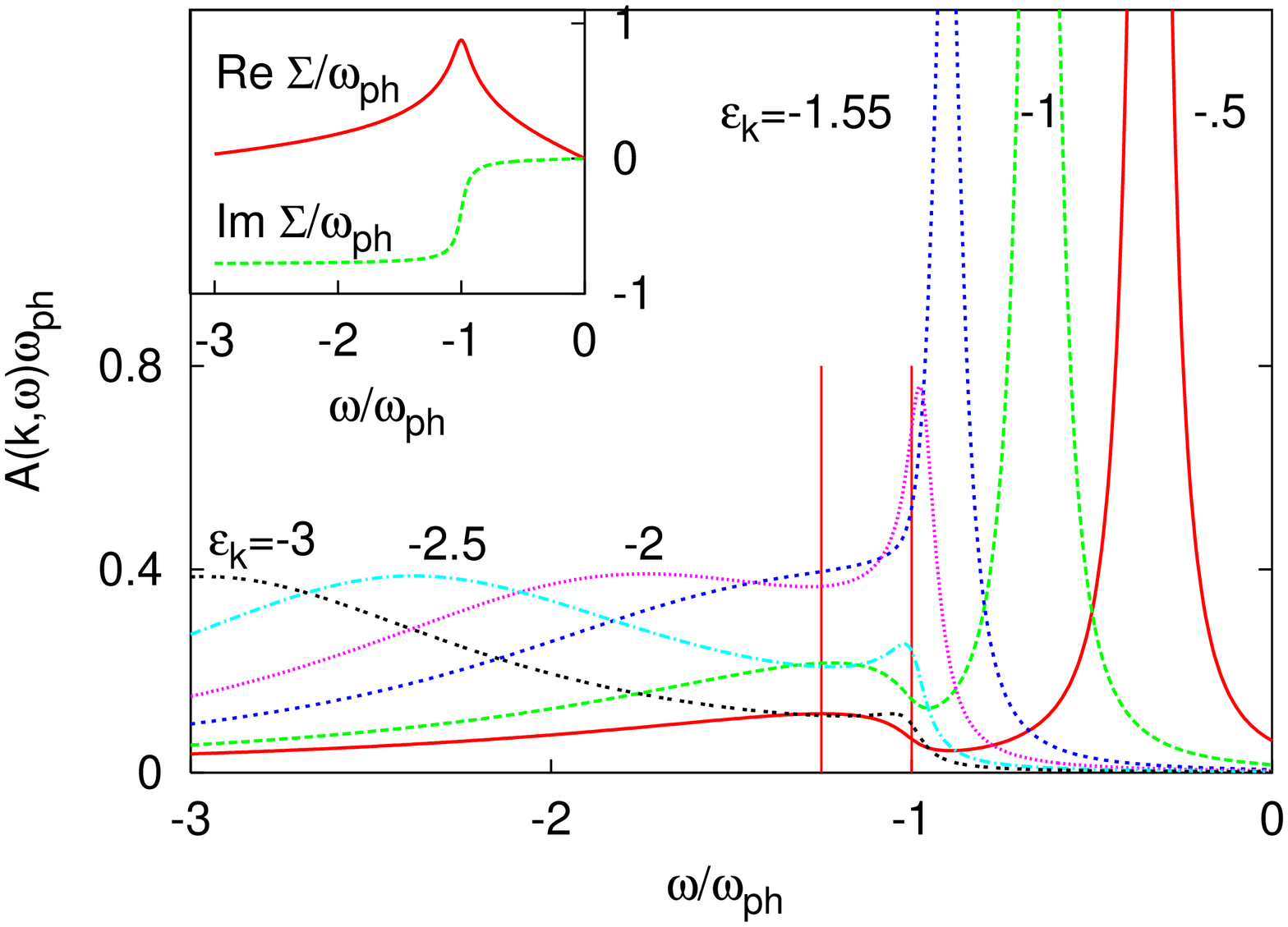}}}}
\hskip-1.0cm
{\rotatebox{-90}{\resizebox{!}{9.0cm}{\includegraphics {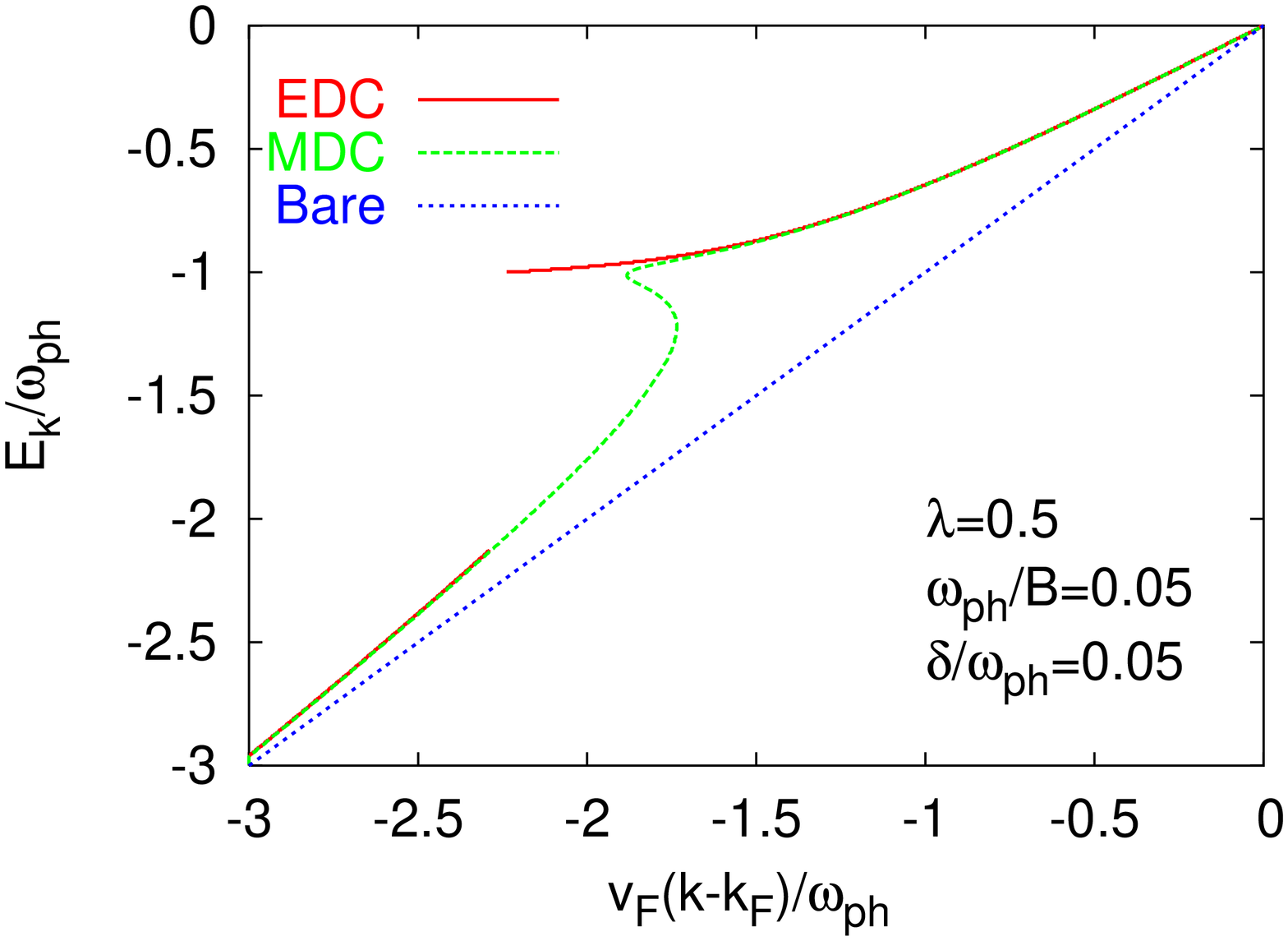}}}}
\caption[]{\label{fig:n1}The spectral function (left), calculated
for different values of the bare energy $\varepsilon_{\bf k}$ (in units of 
$\omega_\mathrm{ph}$), and the dispersion (right), comparing EDC and MDC results 
with the bare dispersion. The inset in the left figure shows the real and 
imaginary part of $\Sigma$. All results are obtained for the model in Eqs. 
(\ref{eq:m4},\ref{eq:n1}) and they have been given a Lorentzian broadening 
(FWHM) of $2\delta$. The parameters are $\lambda=0.5$, $\omega_\mathrm{ph}/B=0.05$ 
and $\delta/\omega_\mathrm{ph}=0.05$. 
}
\end{figure}

Results for $\Sigma$ are shown in the inset of Fig.~\ref{fig:n1}. 
At $T=0$, a hole with a binding energy larger than the phonon frequency
$\omega_\mathrm{ph}$ can excite a phonon and simultaneously be scattered.
$|{\rm Im} \Sigma|$ is therefore large for $\omega<-\omega_\mathrm{ph}$. If the
system is ($s$-wave) superconducting, scattering is only possible for binding  
energies larger than $\omega_\mathrm{ph}+\Delta$, where $\Delta$ is the gap, 
since the scattered electron has at least the binding energy $\Delta$ 
\cite{Scalapino1969}.  Related to the abrupt onset of Im $\Sigma$ there 
is a logarithmic singularity in Re $\Sigma$. The left part of Fig.~\ref{fig:n1} shows the spectral function
\begin{equation}\label{eq:n5}
A({\bf k},\omega)=\frac{1}{\pi}{|{\rm Im }\Sigma({\bf k},\omega)|\over
\lbrack \omega-\varepsilon_{\bf k}- {\rm Re}\Sigma({\bf k},\omega)\rbrack^2
+\lbrack {\rm Im }\Sigma({\bf k},\omega) \rbrack^2 }.
\end{equation}  
For simplicity, in Fig.~\ref{fig:n1} we have assumed a linear dispersion 
over the energy range of interest, $\varepsilon_{\bf k}
=v_\mathrm{F}(k-k_\mathrm{F})$, where $v_\mathrm{F}$ and $k_\mathrm{F}$ are the Fermi velocity and wave 
vector, respectively. $A({\bf k},\omega)$ shows a narrow peak at $E_{\bf k}
=\varepsilon_{\bf k}/(1+\lambda)$ if  $|E_{\bf k}| \ll \omega_\mathrm{ph}$ and
it has a broad peak at $E_{\bf k}=\varepsilon_{\bf k}$ if $|\varepsilon_{\bf k}| 
\gg \omega_\mathrm{ph}$. The electron-phonon coupling therefore leads to a 
change of slope
\begin{equation}\label{eq:n5a}
{|dE_{\bf k}/d{\bf k}|_{E_{\bf k}=0
}\over 
|dE_{\bf k}/d{\bf k}|_{|E_{\bf k}
|\gg\omega_\mathrm{ph}}}=
{1\over 1+\lambda}.
\end{equation}
We also define the quasiparticle strength 
\begin{equation}\label{eq:n8} 
Z({\bf k})={1\over 1-(\partial \Sigma({\bf k},\omega)/\partial \omega)\vert_
{\omega=E_{\bf k}}}={1 \over 1+\lambda},
\end{equation}
where the second equality is only valid for $|E_{\bf k}|\ll \omega_\mathrm{ph}$.

\subsection{Energy and momentum distribution curves}\label{sec:nb}
Figure~\ref{fig:n1} shows energy distribution curves (EDC), where 
$A({\bf k},\omega)$ is shown as a function of $\omega$ for a fixed 
value of ${\bf k}$. For each value of ${\bf k}$ we can determine
the $\omega$ for which $A({\bf k},\omega)$ has its maximum. From this 
we obtain a dispersion relation $E_{\bf k}$ shown by the red curve 
(EDC) in the right part of the figure. From the results for $A({\bf k},
\omega)$ it is clear that the EDC dispersion jumps from $E_{\bf k}/
\omega_\mathrm{ph}\approx -1$ for $\varepsilon_{\bf k}/\omega_\mathrm{ph}\approx -2$ 
to $E_{\bf k}/\omega_\mathrm{ph}\approx -2.1$ for $\varepsilon_{\bf k}=/
\omega_\mathrm{ph}\approx -2.3$. The EDC dispersion is reduced by a factor 
$(1+\lambda)$ close to $E_{\bf k}=0$, as discussed below 
Eq.~(\ref{eq:n5}), and the dispersion is further reduced as $E_{\bf k}$
approaches $-\omega_\mathrm{ph}$.

Alternatively, we can study 
momentum distribution curves (MDC), showing $A({\bf k},\omega)$ as a function 
of ${\bf k}$ for a fixed $\omega$. In particular, we can find the ${\bf k}$ 
for which $A({\bf k},\omega)$ has its maximum ($\omega$ fixed).
This MDC 
dispersion relation is shown in the right part of Fig.~\ref{fig:n1}.
As an 
illustration, the vertical lines in the left part of the figure show the 
energies $\omega/\omega_\mathrm{ph}=-1$ and -1.25. The maximum (among the ${\bf k}$ 
values shown in Fig.~\ref{fig:n1}) of $A({\bf k},\omega/\omega_\mathrm{ph}=-1)$ is 
obtained for $\varepsilon_{\bf k}/\omega_\mathrm{ph}=-2$, while the maximum of 
$A({\bf k},\omega/ \omega_\mathrm{ph}=-1.25)$ is obtained for $\varepsilon_{\bf k}
/\omega_\mathrm{ph}=-1.55$. Over a certain frequency range, an {\it increase} in
$|\omega|$ then leads to a {\it decrease} in $|\varepsilon_{\bf k}|$. The 
result is the S-like shape of the MDC dispersion around $\omega/\omega_\mathrm{ph}=-1$. 
Well away from this energy, the EDC and MDC curves agree for this 
${\bf k}$-independent self-energy. 

$A({\bf k},\omega)$ tends to show a dip at $\omega \approx -\omega_\mathrm{ph}$, in
particular if $0>\varepsilon_{\bf k}>-\omega_\mathrm{ph}$ \cite{Scalapino1969}. 
This is the combined effect of a logarithmic singularity in Re 
$\Sigma({\bf k},\omega)$ at $\omega=-\omega_\mathrm{ph}$, which makes the first 
term in the denominator in Eq.~(\ref{eq:n5}) large, and the sudden onset 
of Im $\Sigma({\bf k},\omega)$ below $\omega=-\omega_\mathrm{ph}$, which makes 
the numerator large. As discussed above, the singularity is shifted to 
$\omega_\mathrm{ph}+\Delta$ in a (s-wave) superconductor. At the same time the 
effect becomes stronger, since weight is piled up at the onset of Im 
$\Sigma({\bf k},\omega)$, leading to a more pronounced structure in 
Re $\Sigma({\bf k},\omega)$. As discussed by Sandvik {\it et al.}
\cite{Sandvik}, this effect is particularly pronounced for an $s$-wave 
superconductor and less strong for a $d$-wave superconductor.

Similarly, the phonon self-energy $\Pi({\bf q},\omega)$ can be calculated 
to lowest order in $g$, where ${\bf q}$ is the phonon wave vector. 
Im $\Pi({\bf q},\omega)$ is of particular interest, since it determines 
the phonon width. For a system of noninteracting electrons and a nondegenerate
phonon mode, Allen \cite{Allen1,Allen2} showed that
\begin{equation}\label{eq:n6} 
2 \ {\rm Im} \ \Pi(\omega_\mathrm{ph})=2\pi \hbar^2\omega_\mathrm{ph}^2 N(0) \lambda,
\end{equation}
where 2 Im $\Pi({\bf q},\omega_\mathrm{ph})$ is the full width at half
maximum (FWHM) and $N(0)$ is the density of states (DOS) per spin at the
Fermi energy. Knowledge of the FWHM of the phonons then gives 
a possibility of estimating $\lambda$.                                        

\subsection{Electron-phonon coupling in the 2d Holstein model}\label{sec:nc}

Two-dimensional (2d) correlated models with EPI are often compared 
with the 2d Holstein model to determine the effects of correlation 
on the EPI. A 2d Holstein model at half-filling with only nearest 
neighbor hopping is unstable to an infinitesimal EPI due to perfect 
nesting. Therefore the comparison is often made to a Holstein model 
with just a single electron at the bottom of the band \cite{Ramsak,Mishchenko}. 
Often a $t$-$J$ model doped with one hole is studied, suggesting         
similarities with a Holstein model with a single electron. The half-filled
Holstein model, however, is of particular interest, since the relevant 
antibonding Cu-O band in the cuprates is close to half-filling.  

\subsubsection{Weak-coupling limit}\label{sec:nc1}

We consider the limit $g\ll t$ and $\omega_\mathrm{ph}\ll t$ for a nearest 
neighbor hopping $t$ with a single electron Holstein model. Using  the self-energy 
in Eq.~(\ref{eq:n2}), we can define an effective electron-phonon 
coupling $\lambda=\lambda_0$ via the quasiparticle weight $Z$ 
[Eq.~(\ref{eq:n8})] or the effective mass 
\begin{equation}\label{eq:n7} 
\left.{d^2E_{\bf k}/ d{\bf k}^2\over d^2\varepsilon_{\bf k}/d{\bf k}^2}\right|_
{{\bf k}=0} = { 1\over 1+\lambda_0},
\end{equation}
where
\begin{equation}\label{eq:n9} 
\lambda_0={g^2\over 4\pi t \omega_{\rm ph}}.
\end{equation}
Both methods give the same $\lambda$ for this model in the weak-coupling
limit \cite{Fehske2000}. Instead of the 
Holstein model with a single electron, we can study the half-filled 
model assuming a constant density of states (DOS) and calculating 
$\Sigma({\bf k},\omega)$ according to Eq.~(\ref{eq:n3}). Defining 
$\lambda$ via the expression for 
$Z(\bf k)$ recovers the $\lambda$ defined in Eq.~(\ref{eq:n3a})
and leads to $\lambda=\pi\lambda_0$, since $N(0)=1/(8t)$. The 
increase in $\lambda$ is partly due to the fact that the self-energy 
at the bottom of the band only has contributions from higher states 
while in the half-filled case there are contributions from both higher 
and lower states and partly due to the DOS at the bottom of the band 
being smaller than the average DOS \cite{Giorgio2006}. 

\subsubsection{Polaron formation}\label{sec:nc2}

Polaron formation has been studied extensively for the Holstein model
\cite{Millis96,Ciuchi97,Benedetti,Hewson,Jong,Capone2003,Mishchenko,Capone2005}.
For noninteracting electrons, polaron formation is often associated with
bipolaron formation \cite{Capone2003,Capone2005}. Since on-site bipolaron
formation is strongly suppressed by the large on-site $U$ relevant for 
cuprates, we focus on polaron formation here. 
We use $Z \to 0$ as the criterion for polaron formation. For the 
nearest neighbor Holstein model with a single electron and $\omega_\mathrm{ph}
=0.0125W$ this was found to happen for $\lambda_c=1.2$ \cite{Mishchenko}, 
where $W$ is the band width and $\lambda$ is here defined as $\lambda=2g^2/W$,
corresponding to the assumption $N(0)=1/W$ in Eq.~(\ref{eq:n3a}).
For the half-filled case with a semi-elliptical DOS and using the 
dynamical mean-field theory (DMFT) \cite{DMFT} is was found that
$\lambda_c=0.33$ \cite{Giorgio2006}, again using $\omega_\mathrm{ph}
=0.0125W$ and putting $N(0)=1/W$ in the definition of  $\lambda$. 
Similar results were found by several other groups
\cite{Hewson,Jong,Capone2003,Capone2005}. As in the weak coupling 
limit, there is a large difference between the single electron
and half-filled cases \cite{Giorgio2006}. 

To understand this difference, we consider polaron formation in the 
adiabatic limit by comparing free electron states with states of perfectly localized electrons \cite{Capone,Giorgio2006}. The energy for free electrons 
is $E_{\rm free} =-4t$ per electron for the single electron case and
$E_{\rm free} =-16t/(3\pi)\approx -1.7t$ for the half-filled case. 
In the localized case, $E_{\rm loc} =-g^2/ \omega_ {\rm ph}$ per electron 
for both cases. For simplicity, we assume that polarons form when 
$|E_{\rm loc}|> |E_{\rm free}|$. This leads to a large $\lambda_c=1$ for 
the single electron case and a much smaller $\lambda_c=0.42$ for the
 half-filled case, in rather good agreement with more accurate calculations. 
The large difference between the two cases is due to the large difference in
kinetic energy per electron. In the single electron case, the electron
is at the bottom of the band and has the maximum (absolute) kinetic  
energy, while in the half-filled case the average kinetic energy
is much smaller. The electron-phonon interaction energy can then  
win more easily and lead to polaron formation.

\section{Experimental results}\label{sec:exp}
 
\subsection{Phonon softening and width}\label{sec:expphon}

\begin{figure}[bt]
{\rotatebox{0}{\resizebox{!}{7.0cm}{\includegraphics {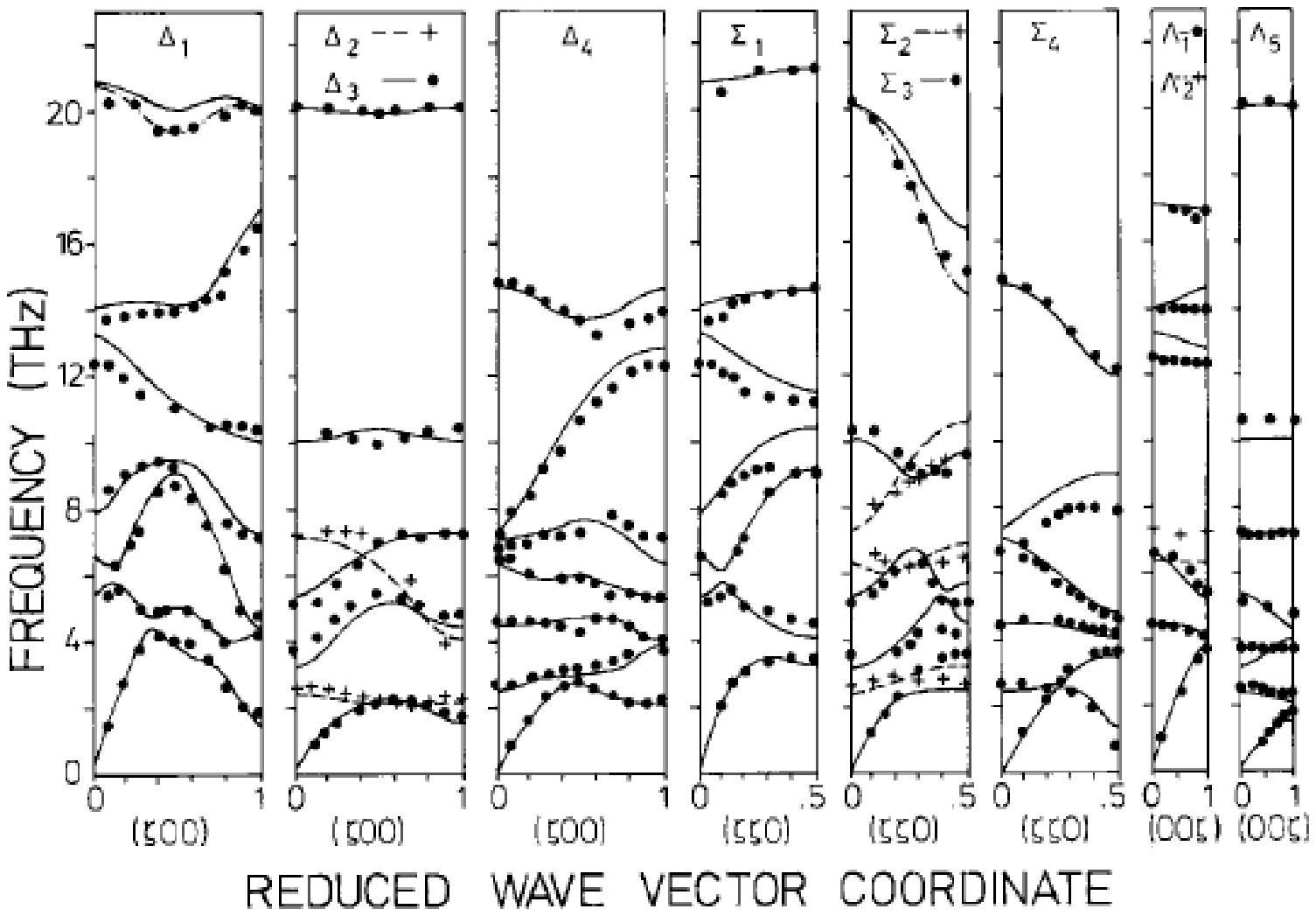}}}}
{\rotatebox{0}{\resizebox{!}{7.0cm}{\includegraphics {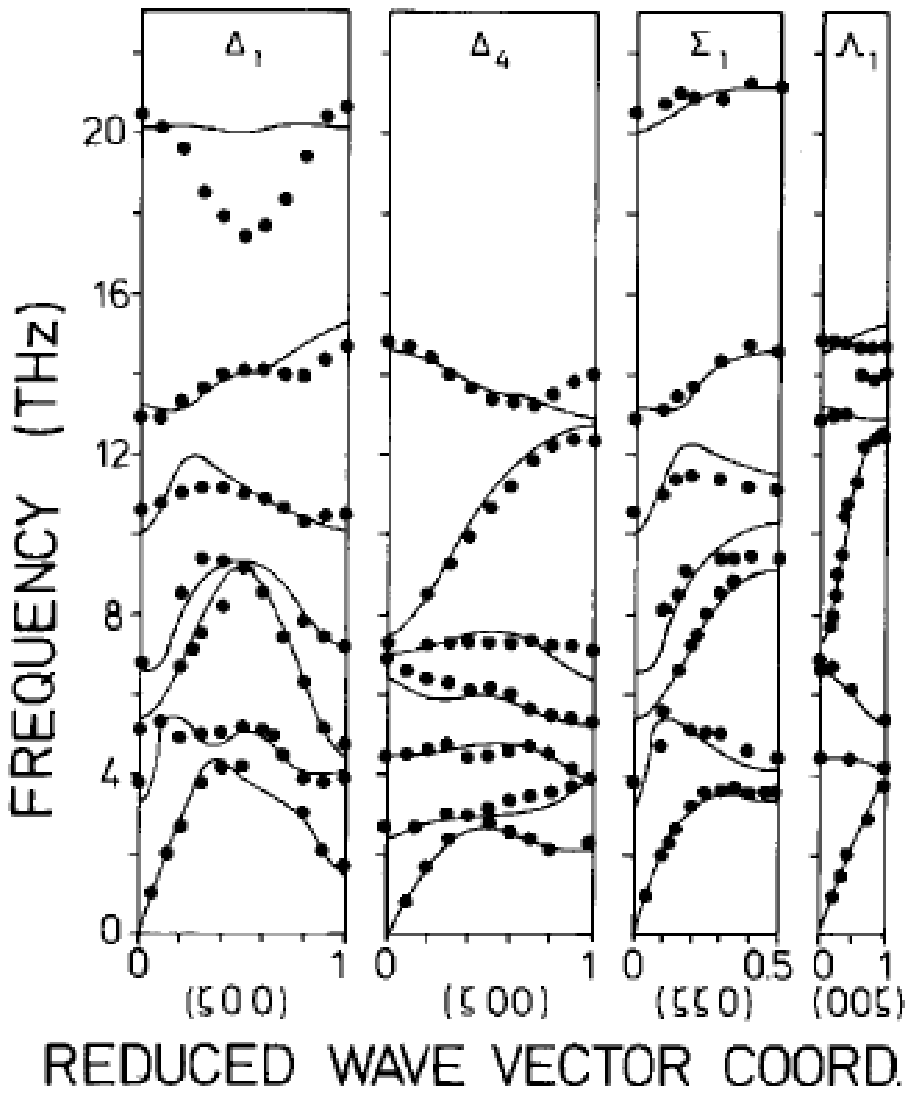}}}}
\caption[]{\label{fig:ep1}Phonon dispersion curves for La$_2$CuO$_4$ (left
hand side) and La$_{1.9}$Sr$_{0.1}$CuO$_4$ (right hand side). The solid curves 
were calculated in a shell model by Chaplot {\it et al.} \cite{Chaplot}. 
The dash-dotted lines were obtained after inclusion of a quadrupolar 
force constant.  The figure illustrates that in the doped compound (right 
hand side) the half-breathing mode in the $\Delta_1$ symmetry at 20 THz 
and several apical oxygen modes in the $\Lambda_1$ symmetry at 12-16 THz 
are softened.
}
\end{figure}

There have been extensive studies of phonons in high-$T_\mathrm{c}$ 
cuprates using neutron scattering. For reviews, see, e.g., 
Pintschovius and Reichardt \cite{Pint1}, Pintschovius \cite{Pint2},
Egami and Billinge \cite{Egami1996} and Egami \cite{Egami2006}. 
Here we focus on results of particular
 interest for the EPI and in particular La$_{2-x}$Sr$_x$CuO$_4$, 
where the most complete results have been obtained. Figure~\ref{fig:ep1} 
shows results of Chaplot {\it et al.} \cite{Chaplot} for undoped (left 
hand side) and doped (right hand side) La$_{2-x}$Sr$_x$CuO$_4$. The 
solid curves have been obtained within a shell model \cite{Chaplot}. 
The shell model calculations give a rather accurate description 
of almost all phonon branches. 

\begin{figure}[bt]
{\rotatebox{0}{\resizebox{!}{4.0cm}{\includegraphics {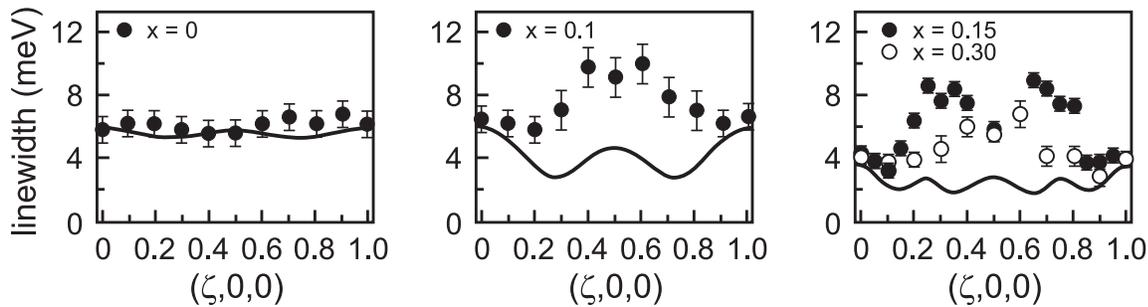}}}}
\caption[]{\label{fig:ep3}Width of the oxygen bond-stretching
phonon for La$_{2-x}$Sr$_x$CuO$_4$ as function of the reduced wave 
vector $(\zeta,0,0)$ and for doping $x=0$ (left), $x=0.1$ (middle) and 
$x=0.15, 0.30$ (right).  The full line shows the experimental 
resolution including focusing effects. The figure illustrates the 
large broadening for the doped system and the small intrinsic 
broadening for the undoped ($x=0$) system (after Pintschovius 
\cite{Pint2}).
}
\end{figure}

One striking exception is the highest mode of $\Delta_1$ symmetry for 
the doped system. This is the so-called half-breathing phonon, which 
is a bond-stretching vibration of the oxygen atoms in the CuO$_2$ plane, 
as shown in Fig.~\ref{fig:ep2}. This phonon is rather well described 
by the shell model for the undoped system. However, doping leads to a 
strong softening half way along the $(1,0,0)$ direction. This softening
is anomalous in the sense that it is not captured by the shell model. 
Anomalies of such bond-stretching phonons have been observed by several 
groups for both high-$T_\mathrm{c}$ cuprates 
\cite{Pint3,Reichardt1,Braden1,Pint4,McQue1,McQue2,Pint5,McQue3}  
and other compounds \cite{Tranquada,Reichardt2,Braden2}. The anomalous 
behaviour of this phonon is also illustrated by the large broadening
in the doped system, as is shown in Fig.~\ref{fig:ep3}. Thus Pintschovius 
and Braden \cite{Pint4} found the FWHM intrinsic broadening for 
${\bf q}=(0.5,0,0)$ to be 1.2~THz=5~meV for 15\% doping.
It is unlikely that this width is caused by disorder, since the width of
the ${\bf q}= (0.5,0.5,0)$ quadrupolar mode (where two of O atoms around 
a Cu move towards Cu and two O atoms move away from Cu) is resolution limited 
\cite{Pint2006}. Both the softening and the width indicate that this phonon 
couples strongly to doped holes. Using the formula [Eq.~(\ref{eq:n6})] of 
Allen \cite{Allen1,Allen2} for noninteracting electrons and the density 
of states $N(0)\sim 0.66$ states per eV and spin \cite{Mattheiss}, the 
electron-phonon coupling can be estimated to be $\lambda \approx 0.1-0.3$ 
for the (half-)breathing phonons. In view of the arguments in Sec.~\ref{sec:is},
this estimate and similar estimates below should be taken with caution.  

Reznik {\it et al.} \cite{Reznik2006} found a strong anomaly for the 
bond-stretching phonon around ${\bf q}=(0.25,0,0)$ for several cuprates.       
This anomaly was particularly large for systems with static stripe order
but it was also seen for systems where stripe order has not been observed,
e.g., superconducting La$_{2-x}$Sr$_x$CuO$_4$. This was interpreted in
terms of a coupling to a dynamic charge inhomogeneity. 

The O$_Z^Z$ phonon of $\Lambda_1$ symmetry at about 17~THz in the undoped
system and with the reduced wave vector (001) shows strong softening under 
doping (see Fig.~\ref{fig:ep1}). As shown to the right in Fig.~\ref{fig:ep2}, 
this phonon mainly involves the movement of apical O.  The softening of 
this phonon was predicted by Falter {\it et al.} \cite{Falter1,Falter2} 
before being observed experimentally. The width of the O$_Z^Z$ phonon is 
16 meV \cite{Pint1,Pint2}, suggesting a very strong coupling to this mode.                                

There has also been a substantial interest in a B$_{1g}$ phonon
involving out-of-plane and out-of-phase vibrations of oxygen atoms
in the CuO$_2$ plane. This phonon has an energy of about 42~meV 
for YBa$_2$Cu$_3$O$_{7-\delta}$. The phonon has been studied using 
both Raman \cite{Macfarlane,Cooper,Cardona1,Cardona2,Cardona3,Devereaux99} 
and neutron scattering \cite{Pyka,Reznik}. This phonon shows an 
interesting change of frequency and width as the compound is 
cooled below $T_\mathrm{c}$ \cite{Macfarlane,Cooper,Cardona1,Pyka,Reznik}, 
and a Fano line shape is observed in Raman scattering 
\cite{Cooper,Cardona1}.  Fitting of the line shape and changes of the
spectrum as the sample is cooled, leads to estimates of the electron-phonon 
coupling in the range $\lambda \sim 0.02-0.06$ \cite{Cardona2,Cardona3,Devereaux99} 
for YBa$_2$Cu$_3$O$_{7-\delta}$, comparable to theoretical estimates 
$\lambda=0.02$ from band structure calculations \cite{Cardona3} in the 
local density approximation. Qualitatively similar but smaller effects 
have also been observed for other phonons \cite{Pyka,Harashina}.

\subsection{Photoemission spectra}\label{sec:exppes}

Photoemission spectra have been extensively reviewed 
\cite{Tohyama,Fink2005,Zhou2006,Yoshida2006}, and here we only show 
a few typical results of particular interest in the context of the EPI.

\begin{figure}[bt]
\centerline{
{\rotatebox{0}{\resizebox{!}{9.0cm}{\includegraphics {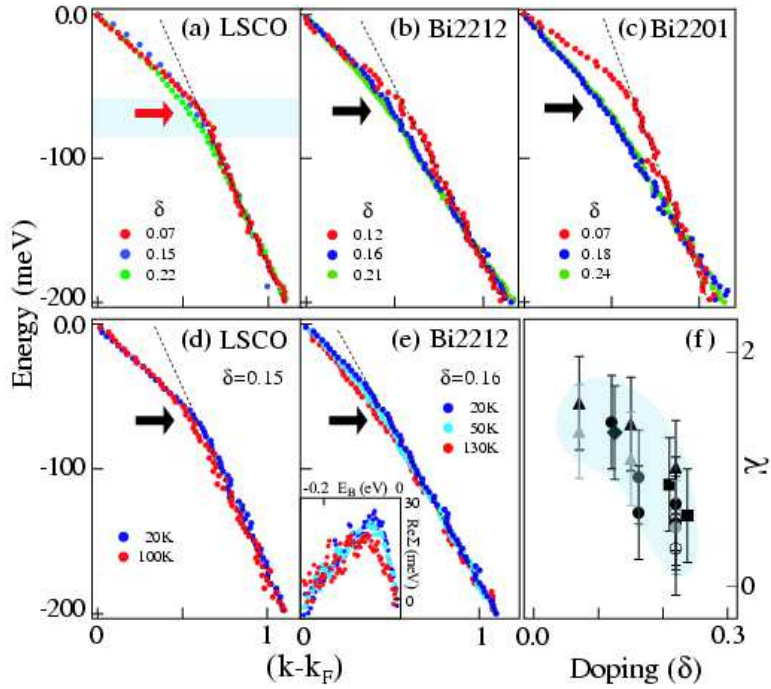}}}}}
\caption[]{\label{fig:el1}Dispersion in the $(0,0)$-$(\pi,\pi)$ direction 
for La$_{2-\delta}$Sr$_{\delta}$CuO$_4$ (LSCO), Bi$_2$Sr$_2$CuO$_{6}$ 
(Bi2201) and Bi$_2$Sr$_2$CaCu$_2$O$_{8}$ (Bi2212) for different dopings 
$\delta$ and temperatures $T$. The results in (a) and (b)
were obtained for $T=20$ K and in (c) for $T=30$ K. The red arrow 
shows the energy of the ${\bf q}=(\pi,0)$ half-breathing phonon and
the black arrows the energies of the kinks. Panel (f) shows 
the change $1+\lambda'$ of the slope at the kink (after Lanzara 
{\it et al.} \cite{Lanzara2001}).
 
}
\end{figure}

Much interest in the EPI has been created by the observation of a kink
in the experimental electron dispersion for several cuprates
\cite{Valla99,Kaminski2000,Bogdanov,Kaminski2001,Johnson2001,Lanzara2001,Gromko2003,Sato2003}.
Some typical results are shown in Fig.~\ref{fig:el1}. Lanzara 
{\it et al.} \cite{Lanzara2001} emphasized that such a kink is found 
for three different families of compounds (LSCO, Bi2201 and Bi2212), 
for different dopings and both below and above $T_\mathrm{c}$. Other groups 
obtained similar results for Bi compounds, but disagreed about whether 
there is a kink above $T_\mathrm{c}$ \cite{Gromko2003,Sato2003} or not 
\cite{Kaminski2001,Johnson2001}. The structure for LSCO, both
below and above $T_\mathrm{c}$, is more pronounced than for B2212 (see 
Fig.~\ref{fig:el1}).
For noninteracting electrons, the ratio of the slopes below and above 
the kink is expected to be given by the dimensionless electron-phonon 
coupling $1+\lambda$ (Eq.~(\ref{eq:n5a})). Figure~\ref{fig:el1}f 
shows the change of slope $1+\lambda'$. Here $\lambda^{'}$ may be different
from $\lambda$, since even the states at large binding energy may not 
show the ``bare'' dispersion. If the theory for noninteracting 
electrons were applicable (Eq.~(\ref{eq:n5a})), this would suggest a 
coupling of the order of $\lambda\sim 1$. 

While the early measurements showed only one kink at about 70~meV, 
later work found several structures at smaller binding energies 
\cite{Zhou2005,Meevasana1}. Estimates of Re $\Sigma({\bf k},\omega)$ 
were extracted from experiment by assuming that the underlying ``bare'' 
dispersion is of second order in $|{\bf k}|-k_\mathrm{F}$ over the range of 
interest. The second derivative of Re $\Sigma({\bf k},\omega)$ for
La$_{2-x}$Sr$_x$CuO$_4$ then has structures at about 40-46 and 58-63~meV
and possibly at 23-29 and 75-85 meV, suggesting that there is coupling 
to bosons at these energies.

The dispersion of the quasiparticle for Bi$_2$Sr$_2$Ca$_{n-1}$Cu$_n$O$_{2n+4}$
away from the nodal direction and, in particular, in the antinodal 
$(\pi,0)$ direction has been extensively studied 
\cite{Norman1997,Kaminski2001,Gromko2003,Sato2003,Kim2003,Cuk2004}. 
For Bi2212 and Bi2223 below $T_\mathrm{c}$ there is a very strong 
structure away from the nodal direction. As discussed in Sec.~\ref{sec:n}, 
if the kink is due to a coupling to a mode at energy $\omega_\mathrm{B}$, 
the kink is expected to appear at roughly $\omega_\mathrm{B}+\Delta$ 
in the superconducting state \cite{Scalapino1969,Norman1998}, where 
$\Delta$ is the gap. This suggests $\omega_\mathrm{B}\approx 40$~meV 
for optimally doped Bi2212. The boson could be the B$_{1g}$ buckling
phonon mode \cite{Cuk2004}, which has roughly the right energy.
Alternatively, the coupling \cite{Kim2003} could be to the 
so-called resonance peak, seen in inelastic neutron scattering in the 
superconducting states of Bi2212 and YBCO \cite{Rossat,Mook,Fong95,Fong99}.
Recently, a broad structure has also been seen at 40-70 meV the inelastic 
neutron scattering spectrum for La$_{2-x}$Sr$_x$CuO$_4$ \cite{Vignolle}.

\subsection{Chemical potential. Polarons}\label{sec:expchem}
\begin{figure}[bt]
\centerline{{\rotatebox{0}{\resizebox{!}{7.0cm}{\includegraphics{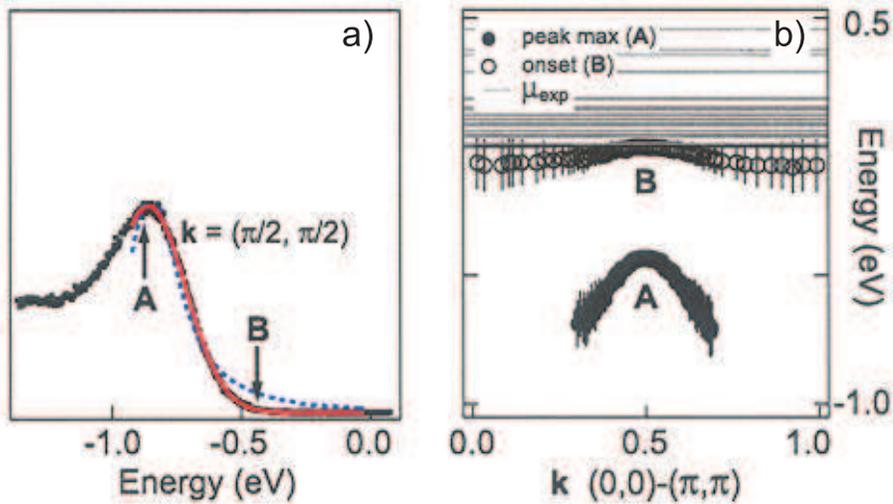}}}}}
\caption[]{\label{fig:c1}a) The ${\bf k}=(\pi/2,\pi/2)$ spectrum of 
undoped Ca$_2$CuO$_2$Cl$_2$ as well as fits to a Gaussian (solid red curve)
and a Lorentzian (dashed blue curve). The maximum of the broad feature is 
denoted by A and its onset by B. b) Dispersion of A and B along the nodal 
direction as well as the different values of the chemical potential $\mu$ 
for a large number of samples (after Shen {\it et al.} \cite{Khyle2004}). 
}
\end{figure}

A long-standing problem has been the position of the chemical potential
in undoped or strongly underdoped cuprates as well as the interpretation 
of the PES spectra for these systems. Figure~\ref{fig:c1}a shows the spectrum 
of undoped Ca$_2$CuO$_2$Cl$_2$ at the top of the band [${\bf k}=(\pi/2,\pi/2)$].
The spectrum has a broad feature centered at A and an onset at B. 
Figure~\ref{fig:c1}b shows the dispersion of these features. 
The dispersion of A (Fig.~\ref{fig:c1}b) agrees well with the extended 
$t$-$J$ model \cite{Tohyama}. It has therefore often been assumed that 
A represents a quasiparticle which is very strongly broadened by some 
unknown mechanism. 

Shen {\it et al.} \cite{Khyle2004} pointed out, however, that  
the peak shape is not Lorentzian, as would be expected from a life-time
broadening, but Gaussian. Even more seriously, a quasiparticle at the 
top of the valence band of an insulator cannot decay into an electron-hole 
pair, and one would expect the width to be small \cite{Khyle2004}.
In Fig.~\ref{fig:c1}b, each horizontal line shows the chemical potential 
of a specific sample. Since the system is an insulator, one might expect 
the chemical potential to be pinned to impurities or defects and therefore
depend on sample preparation. The figure shows, however, that although 
the value of the chemical potential is highly sample dependent, it is always 
at least about 0.45~eV above the peak A\@. Shen {\it et al.} \cite{Khyle2004} 
pointed out that all these puzzling features can be explained by assuming 
polaronic behaviour due to strong coupling to bosons. Peak A is then a
boson side-band, explaining its large width and Gaussian shape, and B 
represents the quasiparticle, having too small a weight to be seen experimentally. 
This explains why the chemical potential is never lower than B in 
Fig.~\ref{fig:c1}b, since this is the top of the valence band in the new
interpretation. It has furthermore been observed that the width of the broad 
peak increases substantially with $T$, providing further support to 
the interpretation in terms of coupling to bosons \cite{Khyle2005}.  

As the doping is increased, a quasiparticle is observed in 
Ca$_{2-x}$Na$_x$CuO$_2$Cl$_2$ for $x=0.1$ \cite{Khyle2004}. The 
quasiparticle weight is small, however, and there is still much 
weight in the energy range of the side band. This suggests that 
the effects of the electron-boson interaction remain important 
for the doped system, although (small) polaronic effects
are not seen anymore. For La$_{2-x}$Sr$_x$CuO$_4$ this is found 
to happen at smaller dopings ($x=0.03$) \cite{Yoshida2003}, but 
it has been proposed that this could be due to phase separation 
\cite{Yoshida2006a}. 

\subsection{Isotope effects}\label{expiso}

Isotope effects provide a good indication for electron-phonon 
interactions. In particular, the interest has focused on the 
superconductivity transition temperature $T_\mathrm{c}$. A review of this 
work is given by Franck \cite{Franck1}. Generally, a small oxygen 
isotope effect was found for compounds having the composition which 
gives the highest $T_\mathrm{c}$ in that family. However, the isotope effect 
is often much larger for systems where $T_\mathrm{c}$ is suppressed by some 
substitutions. 

Khasanov {\it et al.} \cite{Khasanov} observed an isotope effect in the 
penetration depth of nearly optimally doped YBa$_2$Cu$_3$O$_{7-\delta}$, 
using a muon-spin rotation technique, which allows a direct observation 
of the penetration. These results were interpreted in terms of
an isotope effect in the carrier mass due to a strong electron-phonon
interaction violating Migdal's theorem \cite{Khasanov}. 

Gweon {\it et al.} \cite{Gweon} studied the oxygen isotope effect on 
the PES spectra for optimally doped Bi$_2$Sr$_2$CaCu$_2$O$_{8+\delta}$
and found a very large effect, involving shifts as large as 30-40 meV. 
Douglas {\it et al.} \cite{Douglas} in contrast found no large isotope 
effect, but concluded that their measurements are not inconsistent with 
a conventional isotope shift of the order of 3 meV.

\subsection{Scanning tunneling spectroscopy}\label{expstm}

There has recently been much work based on scanning tunneling 
spectroscopy (STM), observing strong spatial modulations and 
a checkerboard structure \cite{Hoffman,McElroy,Hanaguri,STMDavis}. 
Of particular interest here is the work of Lee {\it et al.} 
\cite{STMDavis}, studying  Bi$_2$Sr$_2$CaCu$_2$O$_{8+\delta}$.
From the second derivative of the tunneling current d$^2$I/dV$^2$ they 
determined the spatially dependent gap $\Delta({\bf r})$ and the energy 
of a bosonic mode $\Omega({\bf r})$ appearing in the spectrum. The mode 
energy showed the isotope effect expected for phonons involving primarily 
O atoms. Based on this and on the doping independence of $\Omega({\bf r})$, 
it was concluded that the mode is an O phonon. Its average energy was 
found to be 52 meV, with a substantial spatial variation. They found 
anticorrelation between $\Omega({\bf r})$ and $\Delta({\bf r})$.   
They concluded that the results could be due to i) a heterogeneity
of the frequency and coupling constants of a pairing-related phonon
causing a disorder in $\Delta({\bf r})$, ii) inelastic scattering of 
tunneling electrons by phonons unrelated to superconductivity 
\cite{Sigrist} or iii) a competing phase coupling to phonons and 
causing the anticorrelation between $\Omega({\bf r})$ and $\Delta({\bf r})$.

\section{Interplay between Coulomb and electron-phonon 
interactions}\label{sec:i}           

In systems where the Coulomb repulsion $U$ is important, the effects of
the electron-phonon interaction can be strongly influenced. This can 
easily be seen for the Holstein-Hubbard model [Eqs.~(\ref{eq:m2}, 
\ref{eq:m4})]. The Hamiltonian can be transformed so that the phonons
couple to the deviation of the site occupancies from their average.
If $U$ is small, the number of electrons on a given site fluctuates 
strongly, and there are substantial deviations from the average, 
even for a half-filled system. However, as $U$ is increased, the 
fluctuations are reduced. For $U$ large and close to half-filling, 
most sites then have exactly one electron. For these sites there is 
no electron-phonon coupling. This suggests that the effects of the 
electron-phonon coupling are reduced as $U$ is increased. The 
problem is, however, substantially more complicated, as discussed 
below. For the cuprates, phonon frequencies are typically smaller than
electronic energies. We therefore do not discuss the antiadiabtic 
limit, where phonon frequencies are much larger than electronic 
energies and quite different effects can be found 
\cite{Giorgio2006PRB,Capone2005}.  

\subsection{Sum rules}\label{sec:is}

We first discuss the interplay between Coulomb and electron-phonon 
interactions in terms of sum-rules for the imaginary parts of 
the electron and phonon self-energies. We consider the $t$-$J$ 
model [Eq.~(\ref{eq:m1})] and the electron-phonon coupling
in Eq.~(\ref{eq:m3}) for ${\bf k}$ independent coupling $g({\bf k},
{\bf q})=g({\bf q})$, which results in an on-site coupling. 

\subsubsection{Phonon self-energy}\label{sec:is1}
We first consider the phonon self-energy $\Pi({\bf q},\omega)$. This 
can be written as 
\begin{equation}\label{eq:is2}
\Pi({\bf q},\omega)={(g_{\bf q}^2/N)\chi({\bf q},\omega) \over
 1 + (g_{\bf q}^2/N)\chi({\bf q},\omega)D_0(
{\bf q},\omega)},
\end{equation}
where $\chi({\bf q},\omega)$ is the charge-charge response function and $D_0({\bf q},\omega)$ is the free phonon Green's function.
Khaliullin and Horsch \cite{Horsch1} showed that this
function satisfies a sum rule (at $T=0$) 
\begin{equation}\label{eq:is3}
{1\over \pi N}\sum_{{\bf q}\ne 0}\int_{-\infty}^{\infty} |{\rm Im}\chi(
{\bf q},\omega)|d\omega=2\delta(1-\delta)N,
\end{equation}
where $\delta$ is the doping. This result is suppressed by a factor 
$2\delta(1-\delta)$ compared with the result for noninteracting electrons 
in a half-filled band. Since $\chi({\bf q},\omega)$ becomes small for 
small $\delta$, the denominator in Eq.~(\ref{eq:is2}) is not very important, 
and the sum-rule in Eq.~(\ref{eq:is3}) also applies approximately to 
$\Pi({\bf q},\omega)/g_{\bf q}^2$ \cite{Rosch2}
\begin{equation}\label{eq:is4}
{1\over \pi N}\sum_{{\bf q}\ne 0}{1\over g_{\bf q}^2}
\int_{-\infty}^{\infty} |{\rm Im}\Pi(
{\bf q},\omega)|d\omega \approx 2\delta(1-\delta),
\end{equation}
Since typically $\delta \sim 0.1$, this implies that the softening and 
width of a phonon due to the creation of electron-hole pairs is drastically
reduced. A formula of Allen \cite{Allen1, Allen2} is often used to estimate 
the electron-phonon coupling $\lambda$ from the phonon width 2 Im $\Pi({\bf q},
\omega)$. This formula is derived for noninteracting electrons and it neglects 
the strong reduction of $\Pi$ in Eq.~(\ref{eq:is4}). Its use for high-$T_\mathrm{c}$ 
cuprates may therefore substantially underestimate $\lambda$.

\subsubsection{Electron self-energy}\label{sec:is2}
\begin{figure}[bt]
{{\rotatebox{0}{\resizebox{!}{5.2cm}{\includegraphics{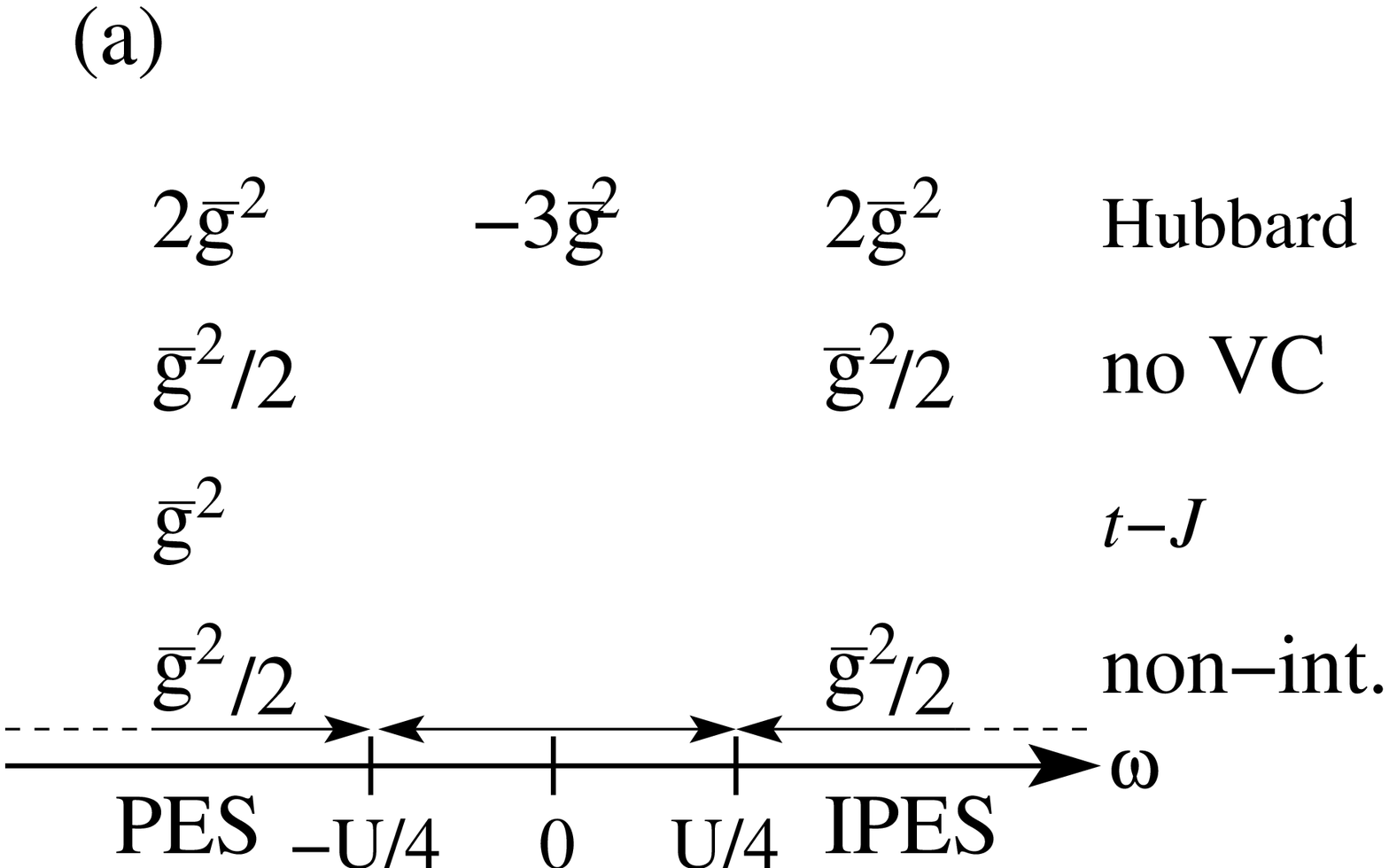}}}}}
{{\rotatebox{0}{\resizebox{!}{5.6cm}{\includegraphics{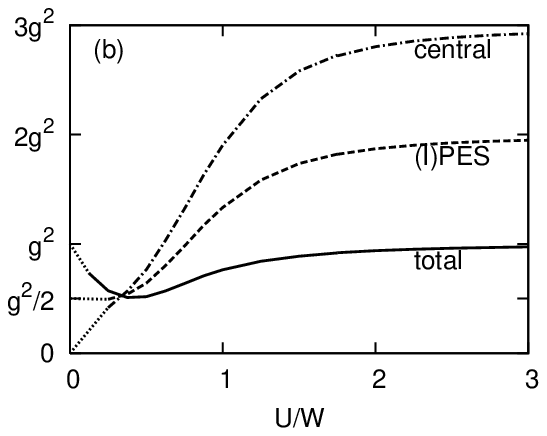}}}}}
\caption[]{\label{fig:is1}
a) Weights obtained by integrating $\mathrm{Im}\ \Sigma_{\rm ep}({\bf k},
\omega-i0^+)/\pi$ over the indicated frequency intervals
for the half-filled Hubbard and undoped $t$-$J$ models. Also
shown are the results for the Hubbard model without vertex
corrections (no VC) and the lowest-order (${\bf k}$-averaged)
result for the $U=0$ Hubbard model (non-int.). For the $t$-$J$
model, the photoemission spectrum has been shifted by -$U/2$
and for the $U=0$ Hubbard model the PES and inverse PES (IPES) 
spectra have been shifted by -$U/2$ and $U/2$, respectively. 
b) $U$ dependence of the absolute value of the total and partial sum 
rules for Im $\Sigma_{\rm ep}$ using DMFT with $\lambda=0.0025$. 
The dotted lines indicate the expected small-$U$ behaviour. $W$
is the $U=0$ band width (after R\"osch {\it et al.} 
\cite{Oliververtex}).
}
\end{figure}

To derive a similar sum rule for the electron self-energy, we define a 
Green's function
\begin{equation}\label{eq:is5}
G({\bf k},z)= {a_{\bf k}\over z -\varepsilon_{\bf k}- \Sigma({\bf k},z)},
\end{equation}
where $a_{\bf k}$ is a weight and $\Sigma({\bf k},z)$ is the electron 
self-energy. The $z$-independent part of $\Sigma$ is included
in the energy $\varepsilon_{\bf k}$, so that $\Sigma({\bf k},z)
\sim b_{\bf k}/z$ for large $z$. By studying the large $z$ behaviour 
of $G({\bf k},z)$ and $\Sigma({\bf k},z)$, R\"osch and Gunnarsson 
\cite{Rosch2} related a sum rule for $\Sigma({\bf k},z)$ to moments 
of the electron-phonon interaction part of the Hamiltonian.                
For the $t$-$J$ model and $\delta=0$, the sum rule takes a 
very simple form 
\begin{equation}\label{eq:is6}
{1\over \pi}\int_{-\infty}^{0}{\rm Im}\Sigma_\mathrm{ep}({\bf k},
\omega-i0^{+})d\omega={1\over N}\sum_{\bf q} |g_{\bf q}|^2\equiv \bar g^2,
\end{equation}
where $\Sigma_\mathrm{ep}$ is the difference in 
self-energy for the system with and without EPI. Since the inverse 
photoemission (IPES) spectrum has zero weight for $\delta=0$,
the integration extends only to $\omega=0$. This result is shown in 
Fig.~\ref{fig:is1}a. For noninteracting electrons and to lowest order 
(in $g_{\bf q}^2$), a similar sum rule for $|{\rm Im} \Sigma_\mathrm{ep}({\bf k}, 
\omega-i0^{+})|$ (averaging over ${\bf k}$) gives the right hand side 
$\bar g^2/2$ for integration up to $\omega=0$ or $\bar g^2$ for 
integration over all energies (see Fig.~\ref{fig:is1}a). Thus the 
effect of the electron-phonon coupling is strongly reduced by the 
small doping for the phonon self-energy but not for the electron 
self-energy, at least not in a sum rule sense. 

The sum-rule in Eq.~(\ref{eq:is6}) can be understood as follows. The 
electron Green's function describes the creation of a hole (or an electron). 
The phonons couple 
strongly to the charge of this hole, even if charge fluctuations are 
strongly suppressed elsewhere in the system. On the other hand, if a 
phonon is created, there is only a small fraction $\delta$ of singlets 
that can respond. As a result, the electron-phonon interaction can be 
expected to appear to be a factor of $1/(c\delta)$ stronger in Re 
$\Sigma({\bf q},\omega)$ than in Im $\Pi({\bf q}, \omega)$ (phonon
width), where $c\sim 2-4$ depends on the assumptions about the 
$\omega$-dependencies of Im $\Sigma$ and Im $\Pi$. 

Although the arguments above show that the sum-rule (\ref{eq:is6}) 
should not go to zero for $\delta \to 0$, the result is, nevertheless, 
nontrivial. The derivation depends on the coupling being on-site and 
we have found no simple result for an off-site coupling. The right 
hand side is independent of ${\bf k}$, $t$ and $J$ and it remains 
proportional to $\bar g^2$ even for large $\bar g$. 

Similar results can be derived for the half-filled Hubbard model in 
the large $U$ limit \cite{Oliververtex}. The PES spectrum is 
expected to be close to the result for the $t$-$J$ model, but the sum 
rule differs by a factor of two, due to the different integrated 
weights of the total (PES and IPES) spectra of the two models 
(see Fig.~\ref{fig:is1}a). Integrating $\mathrm{Im}\ \Sigma_{\rm ep}({\bf k},
\omega-i0^+)/\pi$ over all frequencies for the large $U$ Hubbard model 
gives the sum rule $\bar g^2$. Therefore there is a contribution $-3\bar g^2$
close to $\omega=0$, showing how the EPI reduces a large positive contribution
at $\omega=0$ already present for $\bar g=0$. Going from $U=0$ to large $U$, the 
sum rules for the PES or IPES part increase by a factor of four. This 
is illustrated in Fig.~\ref{fig:is1}b. Interestingly, a substantial 
part of this change has already happened when $U$ is comparable to
the $U=0$ band width $W$ \cite{Oliververtex}.  
The phonon sum rule is further illustrated in Sec.~\ref{sec:ps}.

\begin{figure}[bt]
\centerline{{\rotatebox{0}{\resizebox{!}{3.0cm}{\includegraphics{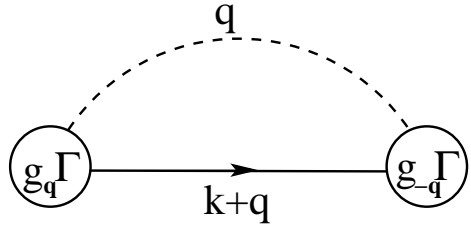}}}}}
\caption[]{\label{fig:ev1}Contribution to $\Sigma_{\rm ep}$. The full and dashed 
lines are electrons and phonon Green's functions, respectively, and the
circle is a vertex correction. 
}
\end{figure}

\subsection{Vertex corrections}\label{sec:iv}

In a diagrammatic language, an important contribution to $\Sigma_{\rm ep}$
is shown in Fig.~\ref{fig:ev1}, expressed in terms of a vertex correction 
$\Gamma$. Figure~\ref{fig:is1}a shows that the neglect of $\Gamma$ (no VC)
in the half-filled large $U$ Hubbard model leads to a large violation of
the sum rules in the previous section. It is therefore interesting to study 
vertex corrections \cite{Kulic1,Castellani,Kulic2,Huang,Cappelluti2004,Koch,Kulic3}.

\begin{figure}[bt]
\centerline{{\rotatebox{0}{\resizebox{!}{5.0cm}{\includegraphics{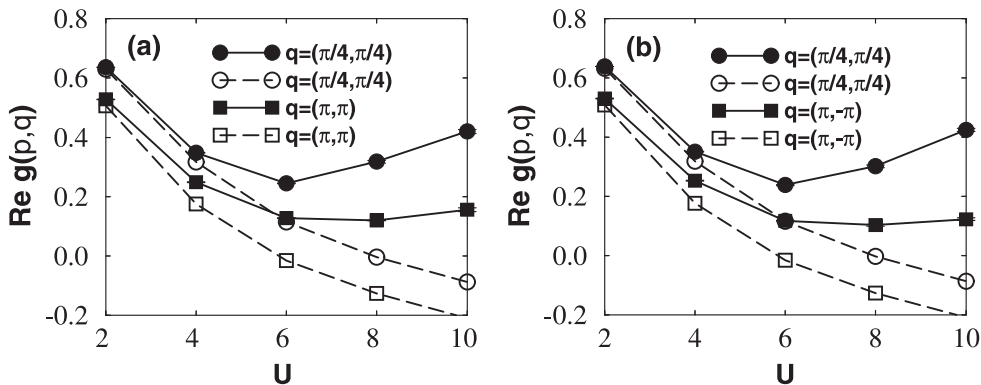}}}}}
\caption[]{\label{fig:ev2}Real part of $g(p, q)$ as a function of $U$
for (a) ${\bf p}=(-\pi,\,0)$ and (b) ${\bf p}=(-\pi/2,\,\pi/2)$.
The value of $\bf q$ is indicated by the shape of the symbol.
The solid circles are Monte Carlo results and the open symbols show
results from perturbation theory (after Huang {\it et al.} \cite{Huang}). 
}
\end{figure}

In particular, Kulic and Zeyher \cite{Kulic1,Kulic2} have argued that in
the large $U$ limit, vertex corrections favor small angle scattering over 
back scattering. This would have the important effect of favoring $d$-wave 
superconductivity. Huang {\it et al.} \cite{Huang}
performed a determinantal quantum Monte-Carlo (QMC) calculation 
\cite{BSS} for  the Holstein-Hubbard model for a $8\times 8$ cluster 
for $T=0.5$, where the hopping integral was put $t=1$ and the band 
width is $W=8$. The occupancy was $n=0.88$. They defined an effective coupling 
\begin{equation}\label{eq:is8}
g(p,q)=\Gamma(p,q) \sqrt{Z(p)Z(p+q)},
\end{equation}
where $\Gamma(p,q)$ is the vertex correction, $p$ and $p+q$ stand 
for the momenta and imaginary frequencies of the incoming and outgoing
electrons, respectively, and $Z$ is the quasi-particle weight. The 
frequencies were put at their minimum value $\pi T$. Results are 
shown in Fig.~\ref{fig:ev2}. It illustrates how $g(p,q)$ is reduced 
relative to its $U=0$ value for all parameters shown, due to the 
reduction of the quasiparticle weights $Z$. Small angle scattering 
$[{\bf q}=(\pi/4,\pi/4)]$, however, is favored over back scattering 
$[{\bf q}=(\pi,\pi)]$ for large values of $U$, which helps $d$-wave 
superconductivity. Some questions have been raised related to the 
large value of $T$ which has to be used in the QMC calculations 
\cite{Cappelluti2004,Koch}.

\subsection{Effects due to antiferromagnetic correlations}\label{sec:ia}

Equation~(\ref{eq:is6}) shows that Im $\Sigma_{\rm ep}({\bf k}, \omega)$ 
is not suppressed by correlation effects in a sum rule sense. The 
sum rule does not tell us, however, how the contributions are 
distributed in frequency. Correlation effects tend to reduce the 
dispersion and provide more low-lying excitations to which a 
quasi-particle could couple. This tends to increase Re $|\Sigma
_{\rm ep}({\bf k},\omega)|$. On the other hand, spectral weight 
is removed from the quasi-particles and shifted away from the 
Fermi energy by correlation effects. This tends to have the 
opposite effect. These two effects need not cancel, as discussed 
below.

\subsubsection{Exact diagonalization}\label{sec:ie}
 
There have been a large number of studies of the Holstein-Hubbard,  
Holstein-$t$-$J$  or related models based on exact diagonalization
\cite{Ranninger92,SchuttlerPRL,Lorenzana94,Fehske94,Fehske95,Fehske96,Sakai,Fehske98,Fehske2000,Fehske2004,Prelovsek2006}. Here we show some typical results.
Zhong and Sch\"uttler \cite{SchuttlerPRL} studied a Holstein-Hubbard
model $H_{\rm Hub}+H_{\rm Hol}$ (Eqs. {\ref{eq:m2}, \ref{eq:m4}) in 
the adiabatic limit ($\omega_\mathrm{ph}=0$) using exact diagonalization 
for an 8-site ($\sqrt{8}\times \sqrt{8}$) cluster with one doped hole, 
i.e., 7 electrons. They found that  the system goes from a delocalized 
state to a polaronic state  for $\lambda\approx 0.2-0.4$ in the 
range $U/t\approx 8-12$. This was compared with a calculation for a 
spin-polarized system with one hole, where the antiferromagnetic spin 
correlations are removed. In this case the  transition to a polaronic
state occurred for $\lambda \approx 0.96$. This was interpreted in 
terms of antiferromagnetic correlations strongly reducing the coupling 
$\lambda$ needed to obtain polaronic behaviour. Similar conclusions 
were obtained for larger cluster and using small but finite phonon 
frequencies by Fehske {\it et al.} \cite{Fehske95,Fehske96,Fehske98}, who 
studied Holstein-$t$-$J$ and Holstein-Hubbard clusters. These calculations
showed that the adiabatic approximation overestimates the tendency 
to (small) polaron formation. The quasiparticle weight and dispersion were 
shown to be reduced as the coupling was increased. B\"auml {\it et al.}
\cite{Fehske98} studied the optical conductivity and argued that the 
midinfrared peak in cuprates is mainly of electronic origin while it
is mainly of polaronic origin in nickelates. 

\subsubsection{Self-consistent Born approximation}\label{sec:iscba}

Ramsak {\it et al.} \cite{Ramsak} used a self-consistent Born 
approximation (SCBA) \cite{SS88,CK90,FM91,GM91,ZL92} for a
 Holstein-$t$-$J$ model, where a hole  is assumed to couple 
to magnons and phonons. The problem is treated in a diagrammatic 
approach, neglecting crossing phonon or magnon lines.  The 
electron-phonon part of the self-energy was written as
\begin{equation}\label{eq:11}
\Sigma_\mathrm{el-ph}({\bf k},\omega)={1\over N}\sum_{\bf q}g_{\bf q}^2
G({\bf k}-{\bf q},\omega-\omega_\mathrm{ph}),
\end{equation}
where $N$ is the number of sites, $g_{\bf q}$ a coupling constant,    
$G({\bf k},\omega)$ the electron Green's function and $\omega_\mathrm{ph}$ 
the phonon frequency. Equation~(\ref{eq:11}) assumes that the 
electron-phonon coupling is not very strong and it neglects 
an indirect contribution, $\Delta \Sigma_m$ \cite{OliverBorn}, via 
the coupling of magnons to the changes of the Green's function due 
to the electron-phonon coupling. To interpret their calculation, 
Ramsak {\it et al.} \cite{Ramsak} argued that the main 
contribution to $\Sigma_\mathrm{el-ph}$ is due to the coherent part
$G_{\rm coh}({\bf k},\omega)=Z({\bf k})/(\omega-\varepsilon_{\bf k})$
of the Green's function, where $Z({\bf k})$ is the quasi-particle 
weight. They furthermore assumed that the quasi-particle energy 
$\varepsilon_{\bf k}$ in the absence of phonons can be parameterized 
in terms of effective masses $m_{\parallel}$ and $m_{\perp}$, 
i.e., $\varepsilon_{\bf k}\approx \tilde \varepsilon_{\bf k}=
k_{\parallel}^2/2m_{\parallel}+ k_{\perp}^2/2m_{\perp}$, where 
$k_{\parallel}$ and $k_{\perp}$ are measured from $(\pi/2,\pi/2)$ 
in the $(\pi/2,\pi/2)\to(0,0)$ and $(\pi/2,\pi/2)\to (\pi,0)$ 
directions, respectively. This gives
\begin{equation}\label{eq:11a}
\Sigma_\mathrm{el-ph}({\bf k},\omega)=4{1\over (2\pi)^2}\int d^2q 
{Z({\bf k-q})g_{\bf q}^2 \over \omega-\omega_\mathrm{ph}-\tilde 
\varepsilon_{\bf k-q}},
\end{equation}
where the factor of four is due to the presence of four hole pockets
at $(\pm \pi/2,\pm \pi/2)$ and the $q$-integrations over the surroundings 
of a hole pocket have been extended to infinity. This leads to a     
simple approximate formula for the enhancement of the mass $m^{\ast}$ 
in the $(\pi/2,\pi/2)\to(0,0)$ direction due to the electron-phonon 
interaction,
\begin{equation}\label{eq:12}
{m^{\ast}\over m_{\parallel}}=(1-4\lambda_0 Z^2    
{\sqrt{m_{\parallel}m_{\perp}}\over m_0})^{-1}\equiv (1+\lambda),
\end{equation}
where $Z$ is approximated by $Z(\pi/2,\pi/2)$, $\lambda_0$ 
[Eq.~(\ref{eq:n9})] is the $\lambda$ obtained for a Holstein model with
just one electron at the bottom of the band and $m_0=1/|2t|$ is the 
corresponding mass. Equation~(\ref{eq:12}) contains an extra factor $Z$ 
\cite{OliverBorn}, obtained from solving the Dyson equation for the 
quasi-particle energy with electron-phonon interaction. As shown in 
Fig.~\ref{fig:11a}, Eq.~(\ref{eq:12}) then agrees rather well with the 
full calculation for a large range of $J/t$ values.  The figure 
illustrates that the use of the coherent part [Eq.~(\ref{eq:11a})] 
is a rather good approximation for large $J/t$, while for small $J/t$ 
it substantially underestimates $\lambda$, primarily due to the neglect 
of $\Delta \Sigma_m$. 

The expression for $m^{\ast}/ m_{\parallel}$ in Eq.~(\ref{eq:12}) is 
reduced by $Z^2$ but enhanced by the large effective mass factor 
$\sqrt{m_{\parallel}m_{\perp}}/m_0$, representing the reduced mobility 
of the carriers due to antiferromagnetic correlations. Thus the 
antiferromagnetic correlations greatly help polaron formation \cite{Ramsak}. 
For large $J/t$, the net effect is an increase of $\lambda/\lambda_0$. 
For $J/t=0.3$, the factor $Z^2\sqrt{m_{\parallel}m_{\perp}}/m_0$ is actually 
smaller than unity, but $\lambda/\lambda_0$ is still enhanced, due to the 
factor of four in Eq.~(\ref{eq:12}), resulting from the four hole pockets.

\begin{figure}[bt]
\centerline{
{\rotatebox{-90}{\resizebox{7.0cm}{!}{\includegraphics {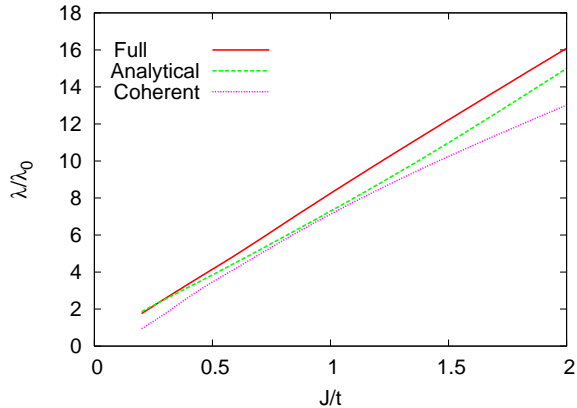}}}}}
\caption[]{\label{fig:11a}$\lambda/\lambda_0$ as a function of $J/t$
for $\omega_\mathrm{ph}/t=0.1$ and $g/t=0.1$. The result of the SCBA 
(full line) is compared with the contribution from the coherent 
part of the Green's function (dashed) and the analytical formula 
(\ref{eq:12}) (after Gunnarsson and R\"osch \cite{OliverBorn}).}
\end{figure}

\subsubsection{Quantum Monte-Carlo calculations}\label{sec:im}

Mishchenko and Nagaosa \cite{Mishchenko} studied an undoped infinite 
Holstein-$t$-$J$ model using a diagrammatic Monte-Carlo method. In 
contrast to the treatment above, they included diagrams with crossing 
phonon propagators and only neglected diagrams with magnon propagators
crossed by phonon or other magnon propagators. Using the parameters 
$\omega_\mathrm{ph}/t=0.1$ and $J/t=0.3$, they found self-trapping for 
$\lambda\approx 1.2$ for an electron at the bottom of the band
in the Holstein model but already for $\lambda\approx 0.4$ in the 
Holstein-$t$-$J$ model \cite{Mishchenko}. It was concluded that  
the antiferromagnetic ground-state and the coupling to magnons 
help the formation of polarons.

\subsubsection{Dynamical mean-field calculations}\label{sec:id}
 
Polaron formation has been extensively studied in the dynamical 
mean-field theory (DMFT) \cite{DMFT}, which becomes exact for 
infinite degeneracy. Cappelluti and Ciuchi \cite{Cappelluti}
developed a method for analytically solving the Holstein-$t$-$J$
model in this limit. They found that antiferromagnetic correlations 
and the electron-phonon interaction mutually reinforce each other.
Cappelluti {\it et al.} \cite{Cappelluti2007} studied the relation 
between polaron formation and structures in the optical conductivity.

The Holstein-Hubbard model in the paramagnetic phase has been 
studied extensively using the DMFT method (P-DMFT) 
\cite{Giorgio2004,Giorgio,Giorgio2006PRB}. It was shown that in   
the P-DMFT the electron-phonon interaction is strongly suppressed
by the Coulomb interaction, in the sense that polaron formation
is suppressed and states close to the Fermi energy are not strongly 
influenced by the EPI. The main effect was found to be a renormalization 
of $U$. As shown in Sec.~\ref{sec:iu}, allowing for antiferromagnetic 
(AF) correlations by using an AF-DMFT method greatly increases the 
importance of the electron-phonon interaction.

\subsection{Differences between different phonons}\label{sec:idd}

\begin{figure}[bt]
\centerline{
{\rotatebox{0}{\resizebox{7.7cm}{!}{\includegraphics {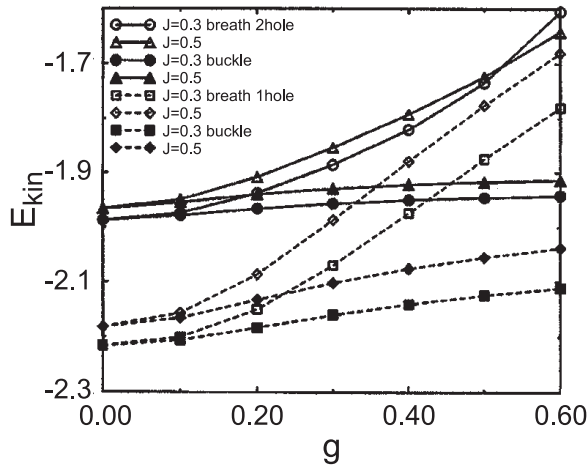}}}}}
\caption[]{\label{fig:l2}Kinetic energy $E_{\rm kin}$ per hole as a function 
of coupling $g$ for a Cu$_8$O$_{16}$ cluster with $\omega_\mathrm{ph}=0.2$. All
energies are in units of the hopping integral $t$. Open (solid) symbols
correspond to the breathing (buckling) phonon. Solid (dashed) lines refer 
to two (one) doped hole. The figure shows how the breathing phonons
reduce hopping more efficiently than buckling phonons (after Sakai 
{\it et al.} \cite{Sakai}).  
}
\end{figure}

Sakai {\it et al.} \cite{Sakai} considered coupling via hopping integrals 
(in a three-band model) to breathing and buckling phonons involving 
in-plane and out-of-plane movements, respectively, of O atom in the 
CuO$_2$ plane. By transforming to a $t$-$J$ model, they obtained an on-site
electron-phonon coupling 
\begin{equation}\label{eq:l3}
H_{BB}=\omega_\mathrm{ph}\sum_{i,{\bf \delta}}b_{i{\bf \delta}}^{\dagger}
b_{i{\bf \delta}}^{\phantom \dagger}+g\sum_{i,{\bf \delta}}
(b_{i,{\bf \delta}}^{\phantom \dagger}
+b_{i,{\bf \delta}}^{\dagger}) (n_i\mp n_{i+{\bf \delta}}),
\end{equation}
where $i$ labels the Cu sites and $\delta=x,y$ differentiates the 
two bond directions. The minus sign between the two occupation numbers 
refers to a breathing phonon and the plus sign to a buckling phonon. 
For a breathing phonon, an O atom between two Cu atoms moves towards 
one Cu atom and thereby away from the other Cu atom, influencing 
Zhang-Rice singlets centered around the two Cu sites in opposite ways, 
as described by the minus sign in Eq.~(\ref{eq:l3}). A buckling 
mode leads to a movement perpendicular to the CuO$_2$ plane of a O atom, 
which influences the two neighboring Cu sites in the same way, giving       
a plus sign in Eq.~(\ref{eq:l3}). The coupling to buckling phonons 
vanishes to linear order for a single perfect CuO$_2$ plane, but it 
is finite for a plane with a static buckling.  

Figure~\ref{fig:l2} shows the kinetic energy $E_{\rm kin}$ as a function 
of the coupling constant $g$ for breathing and buckling phonons. The 
figure illustrates that as $g$ is increased, $|E_{\rm kin}|$ is more 
rapidly reduced for breathing than buckling phonons. If a (local)  
breathing phonon is excited, the potential is lowered on one site 
but increased on a neighboring site. This tends to strongly inhibit 
hopping and reduce $|E_{\rm kin}|$. Exciting a buckling phonon, 
on the other hand, lowers the potential on both sites involved, and 
this inhibits hopping less. This illustrates the importance of what 
type of phonons the electrons couple to. 

\begin{figure}[bt]
\centerline{
{\rotatebox{0}{\resizebox{7.7cm}{!}{\includegraphics {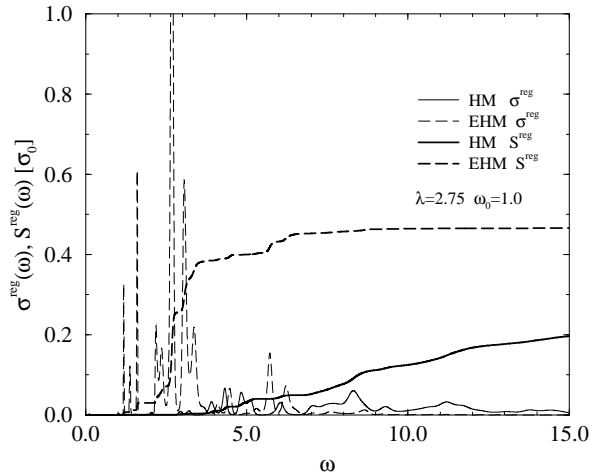}}}}}
\caption[]{\label{fig:ie2}Optical conductivity $\sigma^{\rm reg}(\omega)$
(thin lines) and sum rule $S^{\rm reg}(\omega)$ (thick lines) for 
the 1d Holstein (HM) and extended Holstein (EHM) models in the strong 
coupling limit (after Fehske {\it et al.} \cite{Fehske2000}).
}
\end{figure}

Fehske {\it et al.} \cite{Fehske2000} observed similar effects 
when comparing the one-dimensional Holstein model (HM) to an extended
Holstein model (EHM) with a Fr\"ohlich type long-ranged electron-phonon 
coupling decaying as $1/d^3$ for large $d$, where $d$ is the distance 
between the electron and a phonon. Figure~\ref{fig:ie2} shows the optical 
conductivity $\sigma^{\rm reg}(\omega)$ for the EH and EHM as well as 
the sum rule $S^{\rm reg}(\omega)= \int_0^{\omega}d\omega^{'} 
\sigma^{\rm reg}(\omega^{'})$, where the Drude peak has been removed 
from $\sigma^{\rm reg}(\omega)$. The optical conductivity is determined 
by the current-current correlation function, where the current operator 
for the present models corresponds to the transfer of an electron to a 
neighboring site. In the strong coupling limit, an electron in the HM 
has a large binding energy, $\varepsilon_p$, essentially due to phonons
on the same site. If the electron is moved to a neighboring site by the 
current operator, it looses the energy $\varepsilon_p$ and it leaves 
behind excited phonons with the energy $\varepsilon_p$. Therefore, 
$\sigma(\omega)$ tends to have a broad peak centered at $2\varepsilon_p$,
which is at $\omega=11$ in Fig.~\ref{fig:ie2} \cite{Fehske2000}. Due to 
the long-range of nature of the phonons in the EHM, phonons relatively 
far away from the electron are excited in the EHM, even in the 
strong-coupling limit. Moving an electron to a neighboring site 
then costs much less energy, and $\sigma(\omega)$ has a peak at a much 
smaller energy \cite{Fehske2000}. Coupling to breathing phonons should 
instead shift the peak in $\sigma(\omega)$ to higher energies than in 
the Holstein model for a given $\lambda$. 

There is a substantial coupling to apical oxygen phonons 
\cite{Pint1,Oliver2005}, in particular to modes with a small 
${\bf q}$-vector parallel (${\bf q} _{\parallel}$) to the CuO$_2$ 
plane. These phonons should have a similar effect as the phonons 
in the EHM model above, while modes with $|{\bf q}_{\parallel}|
\sim \pi/a$ should be more similar to a breathing phonon.

\subsection{Effects of Coulomb interaction}\label{sec:iu}
\begin{figure}[bt]
\centerline{
{\rotatebox{-0}{\resizebox{10.0cm}{!}{\includegraphics {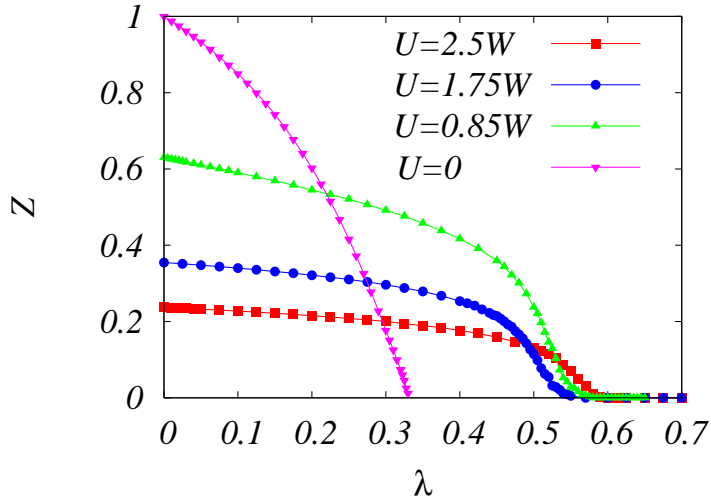}}}}}
\caption[]{\label{fig:id1}$Z$ as a function of $\lambda$ for different 
$U$ and for $\omega_0=0.0125W$, where $W$ is the band width. 
The figure shows how the Coulomb interaction moderately suppresses 
polaron formation ($Z \to 0$) (after Sangiovanni {\it et al.} 
\cite{Giorgio2006}).}
\end{figure}

Most of the work above used the Holstein-$t$-$J$ or large $U$
Holstein-Hubbard model and focused on the effects of antiferromagnetic 
(AF) correlations. Here we focus on the Holstein-Hubbard model, 
which allows a continues increase of $U$ and also to study other
effects of $U$.

In Sec.~\ref{sec:id}, it was found that treating the paramagnetic 
state in DMFT (P-DMFT), leads to a strongly suppressed EPI. Due to the 
nature of P-DMFT, antiferromagnetic (AF) correlations are suppressed, 
which have been found to be important for the EPI (see Sec.~\ref{sec:ia}). 
This suggests that it would be interesting to apply an AF-DMFT method, 
where an AF state is allowed. A second reason for this is that the 
half-filled Holstein-Hubbard model must be an insulator for large $U$. 
In the P-DMFT this can only happen via $Z\to 0$. From Eq.~(\ref{eq:12})
it follows that this strongly suppresses the EPI, at least in the 
weak-coupling limit. In the AF-DMFT, on the other hand, it is possible 
to have an insulating state with $Z>0$. A third reason is to notice 
that P-DMFT is equivalent to solving an Anderson impurity model (with a 
self-consistent host). We consider the electron Green's function, describing,
for instance, the removal of an electron in photoemission. We focus 
on the corresponding final states close to the Fermi energy, i.e., in
the Kondo resonance. These states have essentially the same occupancy 
of the local level as the ground-state, since the electron removed in 
the photoemission process is replaced by an electron hopping in from 
the host \cite{Schonhammer}. Actually, in the limit of infinite orbital 
degeneracy and an infinite $U$, the occupancy of the local level is 
unchanged \cite{PRB83}. Seen from the phonons, coupling to the net charge 
of the local level, a photoemission process corresponding to the Kondo 
peak then leads to no change. As a result, Holstein phonons have only 
an indirect influence on these states due to a renormalization of the 
parameters \cite{Schonhammer,Giorgio2004}. In the Holstein-Hubbard 
model, however, an electron filling the hole created in photoemission 
comes from another $3d$ level, which also couples to phonons, and in  
general this may be expected to influence the spectrum also close to 
the Fermi energy.

The AF-DMFT method has been applied to the Holstein-Hubbard model
on a Bethe lattice using exact diagonalization for solving the
impurity problem \cite{Giorgio2006}. The results for $Z$ are shown in 
Fig.~\ref{fig:id1}. The results for $U=0$ shows how $Z$ is reduced 
from $Z=1$ for $\lambda=0$ to $Z\approx 0$ for $\lambda=\lambda_c=0.33$.
We use this as the criterion for (small) polaron formation. For $\lambda=0$,
an increase of $U$ leads to a decrease of $Z$. However, $Z$ deceases
more slowly with $\lambda$ for a finite $U$, and polaron formation happens 
at a somewhat larger value $\lambda_c$. Thus the Coulomb interaction 
moderately suppresses polaron formation, at least in AF-DMFT. This is 
also shown in Fig.~\ref{fig:id2}. 

It is important to notice that in the half-filled large $U$ case, there is 
no polaron formation in the ground-state, since $U$ suppresses charge 
fluctuations. The Green's function describes the final state after an 
electron has been removed (in, e.g., photoemission) and $Z \to 0$ 
describes how the corresponding hole localizes due to polaronic effects.  
This is different from the Holstein model, where polaron formation means 
the formation of polarons also in the ground-state.

Macridin {\it et al.} \cite{Macridin1} performed a dynamical cluster 
calculation (DCA) for the Holstein-Hubbard model using a $2\times 2$ 
cluster. Using a different definition of polaron formation,
they found $\lambda_c \gtrsim 0.5$, similar to the result in 
Fig.~\ref{fig:id1}. They emphasized the synergistic 
cooperation between the EPI and AF correlations. As a result they found
that the AF transition temperature at finite doping is enhanced by 
the EPI. Macridin {\it et al.} \cite{Macridin1} and Fu {\it et al.} 
\cite{Honerkamp1} pointed out  that the EPI can contribute to a charge 
density modulation seen experimentally.

\begin{figure}[bt]
\centerline{
{\rotatebox{-0}{\resizebox{11.0cm}{!}{\includegraphics {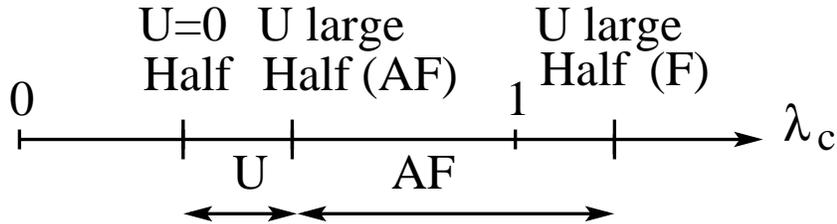}}}}}
\caption[]{\label{fig:id2}Critical value $\lambda_c$ for polaron formation 
($Z \to 0$) in the half-filled Holstein model ($U=0$)  as well as in the large 
$U$ Holstein-Hubbard model in the antiferromagnetic (AF) or ferromagnetic
(F) states. The figure illustrates that AF correlations help the EPI 
(smaller $\lambda$ needed for polaron formation) but that the net effect 
of $U$ is a moderate suppression of the EPI.}
\end{figure}

To see the effects of the antiferromagnetic correlations we compare with 
polaron formation in the ferromagnetic state (F in  Fig.~\ref{fig:id2}),
where antiferromagnetic correlations are completely suppressed. We then 
find that $\lambda_c$ is very large. When the antiferromagnetic correlations 
are reintroduced, $\lambda_c$ is strongly reduced (see Fig.~\ref{fig:id2}), 
meaning that the electron-phonon coupling becomes more efficient, as expected.

Although the antiferromagnetic effects, caused by $U$, strongly reduce 
$\lambda_c$, the net effect of $U$ is still an increase of $\lambda_c$, due 
to other effects of $U$. We may then ask what this is due to. The Green's 
function of the ferromagnetic half-filled state, describes the creation of a  
hole in an otherwise filled spin up band. In the absence of phonons, this hole 
could move completely freely, and if the system has electron-hole symmetry, 
this state is equivalent to a Holstein model with a single electron at the 
bottom of the band. The half-full Holstein model and the single electron 
Holstein model were compared in Sec.~\ref{sec:nc}. It was shown that $\lambda_c$ 
is much larger in the single electron case, because the absolute value of
the hopping energy per electron is much larger. Counterintuitively, 
$\lambda_c$ is therefore larger for the ferromagnetic Holstein model 
than for the half-filled Holstein model because the hopping energy of 
the hole to be localized in the ferromagnetic case is larger, although
the total hopping energy is strongly suppressed.

\begin{figure}[bt]
\centerline{
{\rotatebox{0}{\resizebox{9.0cm}{!}{\includegraphics {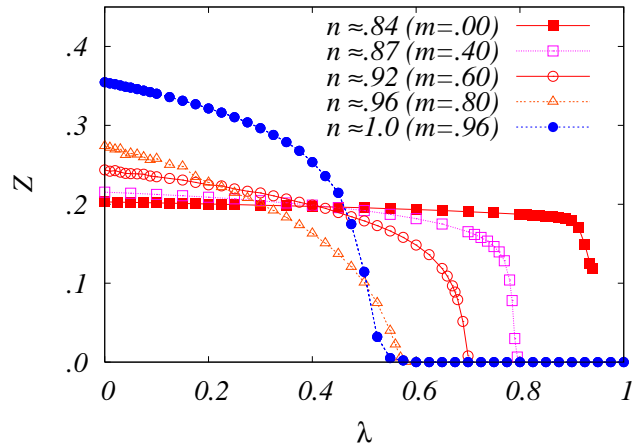}}}}}
\caption[]{\label{fig:id3}$Z$ as a function of $\lambda$ for different 
magnetic moments $m$ and associated fillings $n$ for $U=1.75W$ and 
$\omega_0=0.0125W$. The figure illustrates how the critical $\lambda_c$ 
is increased as the filling is reduced (doping is increased) due to a 
reduction of antiferromagnetic correlations (after Sangiovanni {\it et al.} \cite{Giorgio2006}).}
\end{figure}

The AF-DMFT calculation can also be applied to the doped system.
Figure~\ref{fig:id3} shows $Z$ as a function of $\lambda$ for different
fillings $n$. As the filling is reduced (doping increases) the 
critical $\lambda_c$ for polaron formation increases. The reason
is that AF correlations decrease with increasing doping, which
reduces the effects of the EPI. This is consistent with the
experimental observation \cite{Khyle2004} that polaron formation
is gradually suppressed as the system is doped. In addition to the 
effect discussed here, the calculated strong coupling to apical oxygen
phonons \cite{Oliver2005} becomes more efficiently screened as the system 
is doped, also reducing the tendency to polaron formation.

\subsection{Coupling constants}\label{sec:ic}

Above we have discussed extensively how the EPI is influenced by the
Coulomb interaction and antiferromagnetic correlations by comparing 
the Holstein-$t$-$J$ or Holstein-Hubbard models with the Holstein
model, assuming that the coupling constants remain the same as $U$
is increased. As discussed below, however, the coupling constants 
themselves can change in an essential way.

\begin{figure}[bt]
\centerline{
{\rotatebox{0}{\resizebox{4.0cm}{!}{\includegraphics {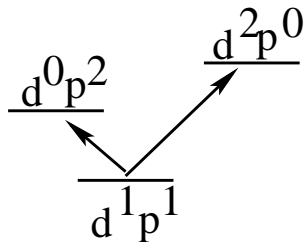}}}}}
\caption[]{\label{fig:icb1}Configurations involved in the formation
of the Zhang-Rice singlet.
}
\end{figure}

\subsubsection{Breathing phonons}\label{sec:icb}

Starting from a three-band model \cite{Emery}, a $t$-$J$ model
with phonons can be derived \cite{Oliverbreathing}. A similar 
derivation can be made for the case when $U=0$ by projecting 
out the oxygen $2p$-states. The EPI is different in the two 
cases, due to the formation of a Zhang-Rice singlet in the large 
$U$ case. The coupling to a singlet involves extra prefactors. 
The main reason for the difference, however, is illustrated in 
Figure~\ref{fig:icb1}. The undoped ground state is nominally a 
$d^{1}$ state in the hole picture. Doping adds a $2p$-hole, 
nominally leading to a $d^1p^1$ configuration. This configuration 
couples to two configurations, $d^0p^2$ and $d^2p^0$. The corresponding 
projection (L\"owdin downfolding) for the $U=0$ case only results in 
the coupling to one configuration \cite{threeband}. The two 
coupling possibilities in the large $U$ case lead to an enhancement 
in the coupling constant squared of the order of three. In addition 
the coupling is screened in the $U=0$ case, which reduces the coupling 
by an additional factor of about two \cite{threeband}. This suggests 
that many-body effects are crucial for the coupling to the breathing 
phonons.

\subsubsection{Apical oxygen phonons}\label{sec:ica}

Neutron scattering experiments \cite{Pint1} and calculations
\cite{Oliver2005} suggest that there is a strong coupling to 
apical oxygen phonons. This strong coupling was found 
\cite{Oliver2005} to depend crucially on the poor screening 
in these systems, in particular for the undoped system, being 
an insulator. This was studied  further by Meevasana {\it et al.} 
\cite{Meevasana1,Meevasana2}. They extracted the effective EPI 
for optimally doped Bi2201 from the structures of the PES 
spectrum in the nodal direction and compared this with the 
experimental loss function in the $c$-direction, obtained from
optical measurements. A rather good agreement was found and it was 
concluded that an essential part of the coupling is due to $c$-axis 
O phonons. For the overdoped sample, they then argued that phonons 
below an energy of the order of 60 meV are screened and do not essentially 
contribute to the EPI. The result is then a reduction of the coupling 
and a stronger emphasis on high-lying phonons, in agreement with 
the EPI extracted from experiments for overdoped Bi2201. 

\section{Phonon spectral function}\label{sec:p}

\subsection{Phonon softening}\label{sec:ps}

\begin{figure}[bt]
{\rotatebox{0}{\resizebox{8.0cm}{!}{\includegraphics {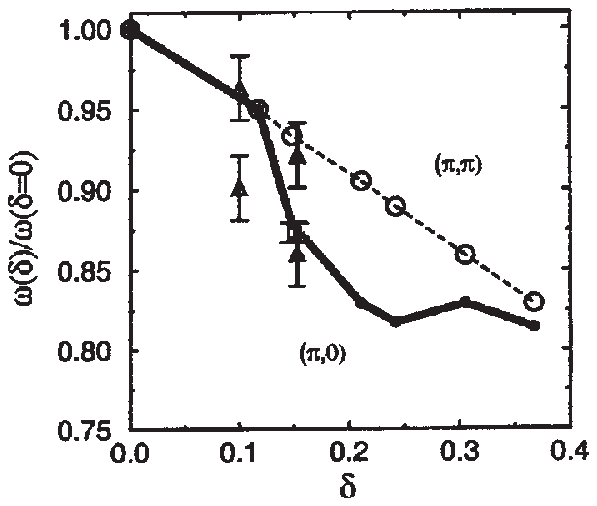}}}}
{\rotatebox{0}{\resizebox{8.0cm}{!}{\includegraphics {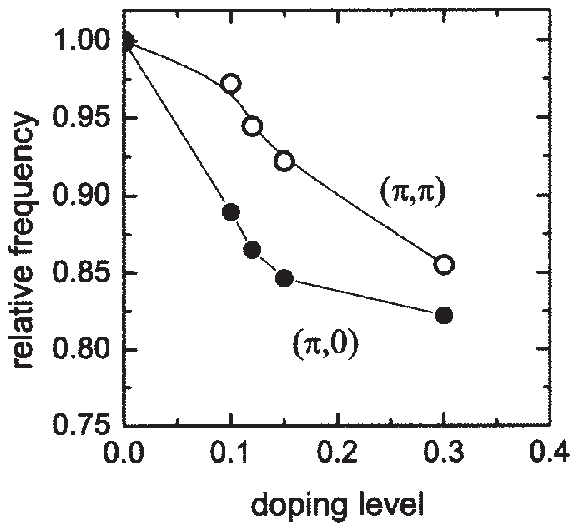}}}}
\caption[]{\label{fig:ps1}Relative softening of the half-breathing
$(\pi,0)$ and breathing phonons $(\pi,\pi)$ as a function of doping $\delta$.
The left figure shows theoretical results of Horsch and Khaliullin
\cite{Horsch4} together with experimental results known at the
time of the calculations and the right figure shows experimental 
results of Pintschovius {\it et al.} \cite{Pint2} (after Pintschovius
\cite{Pint2}).
}
\end{figure}

The softening of the (half-)breathing phonon has been studied by 
several groups
\cite{Becker,Horsch1,Horsch2,Horsch3,Horsch4,Nagaosabreathing,Oliverbreathing}.
Von Szczepanski and Becker \cite{Becker} derived a $t$-$J$ model
with phonons starting from a three-band model. They calculated the 
density response function for the electronic system using
exact diagonalization for a small cluster and from this the phonon 
self-energy and softening. Khalliulin and Horsch
\cite{Horsch1,Horsch2,Horsch3,Horsch4} calculated the density 
response of the $t$-$J$ model using both slave bosons \cite{Horsch1}
and slave fermions \cite{Horsch3}, from which they deduced the 
phonon self-energy. Their calculated relative softening for the
half-breathing $[{\bf q}=(\pi,0)]$ and breathing $[{\bf q}=(\pi,\pi)]$
phonons are compared with experiment in Fig.~\ref{fig:ps1}. 
The theory correctly predicts that for intermediate dopings 
the half-breathing $[{\bf q}=(\pi,0)]$ phonon is softened more than the 
breathing phonon, although the coupling constant is larger for
the breathing phonon [Eq.~(\ref{eq:m4})]. For large dopings
($\delta \approx 0.3$) the softenings of the two modes become
comparable. This was predicted before the experiment had been 
done.   

\begin{figure}[bt]
\centerline{
{\rotatebox{0}{\resizebox{10.0cm}{!}{\includegraphics {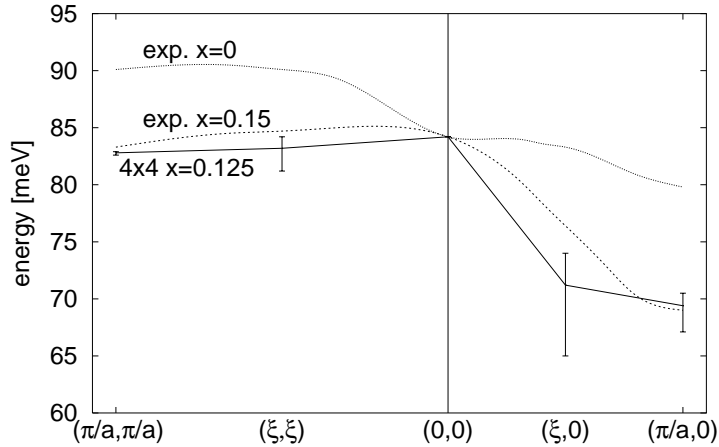}}}}}
\caption[]{\label{fig:ps2}Phonon dispersion in the (1,0) and (1,1)
directions. Experimental results (dotted line) for $x=0$ and $x=0.15$
are shown. Theoretical results (full curve) for $x=0.125$ show the
calculated softening from the experimental $x=0$
results. The average over boundary conditions is shown and the bars
show the spread due to different boundary conditions. There is a strong
softening in the (1,0) direction, while the softening in the (1,1)
direction is weaker at this doping (after R\"osch and Gunnarsson 
\cite{Oliverbreathing}).  }
\end{figure}

R\"osch and Gunnarsson \cite{Oliverbreathing} derived a $t$-$J$ model
with phonons starting from a three-band model. Using input from 
{\it ab initio} calculations they obtained the electron-phonon coupling.
The $t$-$J$ model was solved using exact diagonalization, including 
the phonons in the calculation, and the phonon spectral function was 
calculated. Their calculated dispersion for $\delta=0.125$ is compared
with experimental results in Fig~\ref{fig:ps2}. The dotted curves show 
experimental results and the full curves the calculated softening for 
a $4\times 4$ cluster. The bars show the spread of the results due to
different (periodic, antiperiodic or mixed) boundary conditions. The
figure illustrates that the softening is large for the (1,0) than 
the (1,1) direction for this doping. The softening in the (1,0)
direction is large for $|{\bf q}|\gtrsim \pi/(2a)$. The softening
essentially follows the coupling strength $\sim {\rm sin}^2(q_xa/2)$ 
but is larger at $q_x=\pi/(2a)$ than would be expected from this
argument.            

Sangiovanni {\it et al.} \cite{Giorgio2006} calculated the phonon 
softening within the AF-DMFT theory for the Holstein-Hubbard model.
Figure~\ref{fig:ps3}a shows that there is a large softening for the 
undoped system if $U$ is small. Figure~\ref{fig:ps3}b-c show how the 
softening is much smaller for larger values of $U$, although larger 
values of $\lambda$ were used in Fig.~\ref{fig:ps3}b-c. In particular,
 in Fig.~\ref{fig:ps3}c almost no softening is observed. Finally, 
Fig.~\ref{fig:ps3}d shows results for a doped system. As the doping 
is increased, the softening increases, although $U$ is large. This 
illustrates the sum rule in Eq.~(\ref{eq:is3}) for phonons.

\begin{figure}[bt]
\centerline{
{\rotatebox{0}{\resizebox{10.0cm}{!}{\includegraphics 
{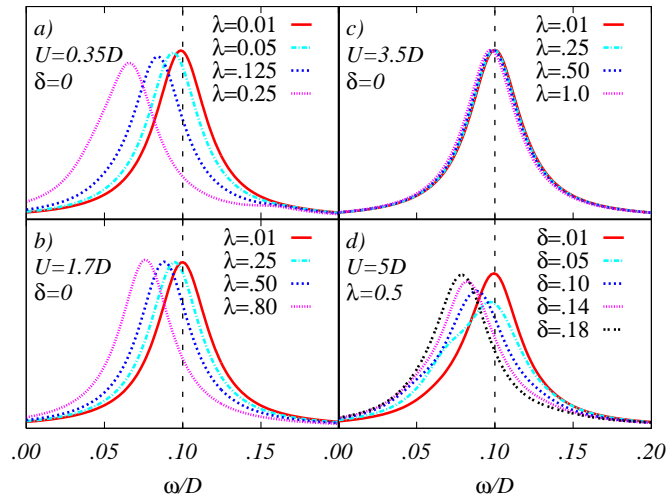}}}}}
\caption[]{\label{fig:ps3}Phonon spectral function for different values of
$\lambda$. The bare phonon frequency is $\omega_{\rm ph}=0.1D$
and a Lorentzian broadening with the full width half
maximum of $0.04 D$ has been introduced, where $D=W/2$ is half the band width.
The figures a-c show how the phonon softening
at half-filling is dramatically suppressed by $U$ and figure
d that the softening increases with doping $\delta$ (after Sangiovanni {\it et al.} \cite{Giorgio2006}).
}
\end{figure}

Falter {\it et al.} \cite{Falter1,Falter2} have developed 
a model for the charge response of cuprates, including the ionic
nature of the system. Based on this model they have calculated 
the softening of several phonons. In particular, they predicted 
the softening of the O$_Z^Z$ phonon (see Fig.~\ref{fig:ep2}) 
\cite{Falter1,Falter2} before it was observed experimentally.  
For La$_{1.85}$Sr$_{0.15}$CuO$_4$ the softening of the O$_Z^Z$ phonon 
is about 30$\%$ and its width is about 4~THz=17~meV \cite{Pint1}. This
large softening and width suggests a strong coupling to doped holes.

\subsection{Phonon width}\label{sec:pw}

Khaliullin and Horsch \cite{Horsch2} have calculated the width of the 
(half-)breathing phonon in the $t$-$J$ model and found that it is very broad 
for ${\bf q}=(\pi,0)$ due to the coupling to a low-lying collective mode 
in the density response function, while the ${\bf q}=(\pi,\pi)$ phonon
is narrower for optimum doping. This is in good agreement with experiment.

It is interesting that LDA calculations \cite{Bohnen} predict the 
frequency of the half-breathing phonon of YBa$_2$Cu$_3$O$_7$ quite 
accurately, while the theoretical width is an order of magnitude 
smaller than the width measured for La$_{2-x}$Sr$_x$CuO$_4$ \cite{Pint1}. 
Since the shift (real part) and width (imaginary part) are both determined
by the phonon self-energy, and since the real and imaginary parts
are related via the Kramers-Kronig relation, this is a surprising result.
This was addressed \cite{threeband} by projecting the three-band model 
onto a one-band model using either the Hartree-Fock (HF) approximation 
or by including many-body effects using the Zhang-Rice method \cite{Zhang} 
to obtain the $t$-$J$ model. The HF approximation shows similarities 
to the LDA approach. As discussed in Sec.~\ref{sec:icb}, many-body 
effects enhance the EPI coupling constants in the derivation of a 
$t$-$J$ model. On the other hand, based on a sum rule in 
Sec.~\ref{sec:is1}, many-body effects lead to a strongly doping 
dependent suppression of the EPI for the phonon self-energy. 
These two effects were shown to roughly cancel for $\delta\approx 0.15$
for the phonon softening \cite{threeband}. Due to many-body 
effects, however, there are particularly many low-lying excitations
which couple strongly to the half-breathing phonon and lead to a
large width for this phonon \cite{threeband,Horsch1,Horsch2,Horsch3,Horsch4}.
This effect is not present in the HF approximation \cite{threeband}.

\section{Electron spectral function}\label{sec:e}

\subsection{Polaronic behaviour in the undoped system}\label{sec:ep}

\subsubsection{Quasiparticle weight in the absence of phonons}
\label{sec:epq}

In Sec.~\ref{sec:expchem} we presented the arguments of Shen {\it et al.}
\cite{Khyle2004} that undoped cuprates show polarons. An essential
part of the argument was that the quasiparticle cannot be seen
experimentally, because its weight $Z$ is so strongly reduced by 
the interaction with the phonons. However, one could also imagine 
that coupling to spin fluctuations alone could have this effect,
and there would then be no need to invoke phonons. We therefore 
first discuss whether or not $Z$ is finite for the undoped $t$-$J$ 
and Hubbard models. 

Some approximate calculations for the $t$-$J$ and Hubbard models gave 
$Z=0$ for $\delta=0$ \cite{Sheng,Paramekanti}. Similarly, DMFT calculations 
\cite{DMFT} for the paramagnetic state of the Hubbard model gave a very small 
$Z$ for a small $\delta$ and a large $U$ \cite{Giorgio,Giorgio2006PRB},
although later calculations including antiferromagnetic correlations 
using the AF-DMFT gave a substantial $Z$ \cite{Giorgioanti}.
It is interesting that in the self-consistent Born approximation
the carrier couples to magnons whose energies go to zero for $|{\bf q}| 
\to 0$. There is, however, no infrared singularity, since the coupling 
also goes to zero, and the quasiparticle weight converges to a finite 
number as the system size goes to infinity \cite{GM91}. 

Exact diagonalization calculations for undoped $t$-$J$ clusters 
with 16, 18, 20, 26 and 32 sites obtained finite values for $Z$, 
and there was no sign of $Z$ going to zero with increasing cluster 
size \cite{Dagotto,Leung}. Brunner {\it et al.} \cite{Brunner} 
studied a $24 \times 24$ $t$-$J$ cluster using a loop algorithm 
and extrapolated the results to the thermodynamic limit. Mishchenko 
{\it et al.} \cite{Mishchenko2001} calculated the Green's function 
$G_k(\tau)$ for imaginary times $\tau$  at $k=(\pi/2,\pi/2)$ for a $32 
\times 32$ $t$-$J$ cluster. By using a continuous-time worm 
algorithm \cite{ProSviTup} they could eliminate any systematic 
errors. From the $\tau$ dependence of $G_k(\tau)$ and from 
analytical continuation \cite{Mishchenko2000} they obtained $Z$. 
Both groups \cite{Brunner,Mishchenko2001} found values of $Z$ 
similar to what had been obtained by exact diagonalization 
for small clusters. These results strongly suggest that $Z$ stays 
finite for $\delta=0$, and that therefore the lack of a visible
quasiparticle in ARPES for undoped cuprates is due to the 
electron-phonon coupling.         

\subsubsection{Effects of phonons}\label{sec:epp}

\begin{figure}[bt]
\centerline{
{\rotatebox{0}{\resizebox{10.0cm}{!}{\includegraphics {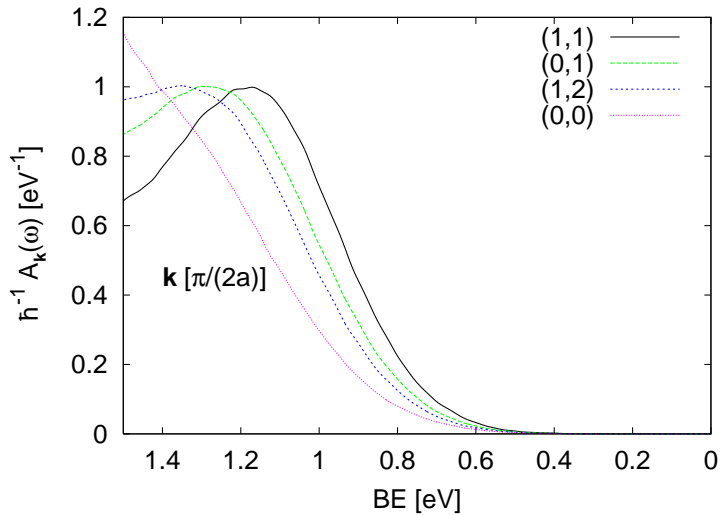}}}}}
\caption[]{\label{fig:epp1} ARPES spectra for the undoped system at 
$T=0$ for different $\bf k$ normalized to the height of the phonon 
side band and as a function of the binding energy (BE) 
(after R\"osch and Gunnarsson \cite{Oliver2005}).
}
\end{figure}

In view of the results in the previous section, it is then 
natural to ask if the electron-phonon coupling is strong 
enough to give polaronic behaviour. This was studied for 
La$_2$CuO$_4$ \cite{Oliver2005} within the $t$-$J$ model 
together with a shell model \cite{Chaplot} for describing the 
phonon eigenvectors. From the eigenvectors one can 
calculate the electrostatic potential acting on a Zhang-Rice 
singlet due to the excitation of a phonon, which provides an
essential part of the coupling. In addition, the phonons modulate
the Cu-O hopping integrals and the energy difference between the 
Cu and O levels in the three-band model used to derive the $t$-$J$
model. This leads to an additional coupling mechanism. Defining 
the coupling as
\begin{equation}\label{eq:ep1}
\lambda=       
2\sum_{{\bf q}\nu}|g_{{\bf q}\nu}|^2/(8t\omega_{{\bf q}\nu}N),
\end{equation}
$\lambda=1.2$ was obtained \cite{Oliver2005}. This is well above
the values $\lambda_c=0.4$ for the Holstein-$t$-$J$ model \cite{Mishchenko}
and $\lambda_c=0.55$ for the Holstein-Hubbard model \cite{Giorgio2006}
giving small polarons.

We find that the dominating coupling for La$_2$CuO$_4$ is due to 
the (half-)breathing phonons, several apical oxygen phonons and 
some low-lying modes involving mainly La and Cu atoms. This is 
supported by inelastic neutron scattering experiments, showing a
large broadening and softening under doping for the (half-)breathing 
and O$_Z^Z$ apical oxygen phonons.  

The resulting $t$-$J$ model with phonons was solved using a method 
based on a statistical sampling of the phonons and exact diagonalization
\cite{Oliver2005,OliverEPJ,Schonhammer}. The results are shown in 
Fig.~\ref{fig:epp1}. 
The binding energy is measured in relation to the quasiparticle, 
which is too small to be seen. The peaks at about -1.1 to -1.3 eV
are phonon side bands due to many unresolved phonon satellites. The
width (0.5 eV) of the side band is close to the experimental result 
(0.47 eV) \cite{Oliver2005}, and also the $T$-dependence is in 
agreement with experimental observations \cite{Oliver2005,Mishchenko2007}. 
The binding energy of the phonon side band is too large 
compared with experiment, which may indicate that the calculation
overestimated the coupling strength. Reducing all coupling constants
by a factor 0.8 (giving $\lambda=0.75$) gives the width 0.4 eV and the binding
energy 0.6 eV, in reasonable agreement with the experimental results
0.47 eV and 0.5 eV, respectively. These results suggest that the 
EPI is sufficiently strong to give the polaronic behaviour seen
experimentally.

\begin{figure}[bt]
\centerline{
{\rotatebox{0}{\resizebox{8.0cm}{!}{\includegraphics {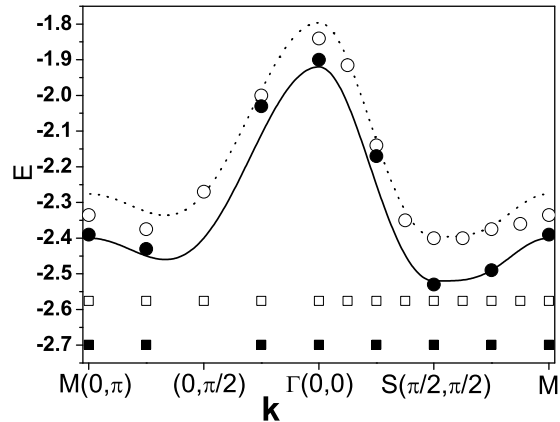}}}}}
\caption[]{\label{fig:epp2}Dispersion of the phonon side band (filled
circles) and the quasiparticle (filled squares) for $\lambda=0.46$ 
as well as the dispersion of a broad peak (open circles) and the  quasiparticle 
(open squares) for $\lambda=0.4$. The two curves show the dispersion 
of a pure $t$-$J$ model (\cite{Horsch89,FM91}) with the energy zeros defined 
appropriately for comparison with $\lambda=0.46$ (full line) and 
$\lambda=0.40$ (dotted line) (after Mishchenko and Nagaosa 
\cite{Mishchenko}). 
 
}
\end{figure}

As discussed in Sec.~\ref{sec:expchem}, the dispersion of the main peak
in ARPES spectra of undoped cuprates can be well described by the 
dispersion of the quasiparticle in an extended $t$-$J$ model with 
up to third nearest neighbor hopping \cite{Tohyama}. However, this 
model cannot explain the large width of the peaks. Shen {\it et al.} 
\cite{Khyle2004} therefore proposed that the peaks are actually 
not quasiparticles but phonon side bands. This then raises the 
question of why the phonon side bands should disperse like the
quasiparticles in a model without phonons. Mishchenko and Nagaosa
\cite{Mishchenko} performed diagrammatic Monte-Carlo calculations 
for a $t$-$J$ model with and without phonons. Their results are shown
in Fig.~\ref{fig:epp2}. For $\lambda=0.46$ there is a well developed 
phonon side band (filled circles) which is found to very closely 
follow the dispersion of the quasiparticle peak for $\lambda=0$ (full line), 
strongly supporting the interpretation of Shen {\it et al.} \cite{Khyle2004}.
A simple explanation of the results of Mishchenko and Nagaosa
\cite{Mishchenko} has been given \cite{OliverEPJ}.
As expected, the weak quasiparticle (filled squares) for $\lambda=0.46$
shows almost no dispersion. For $\lambda=0.4$ the phonon side band 
is less well developed.    

\subsection{Differences between phonons and spin fluctuations}
\label{sec:epd}

The $t$-$J$ model with one hole can be replaced by a model
where the hole couples to magnons, treated as bosons 
\cite{SS88,CK90,FM91,GM91,ZL92}. In this approximation, phonons
and spin fluctuations are treated on the same footing and can
be directly compared. In particular, we can compare the couplings
as defined by Eq.~(\ref{eq:ep1}). The coupling to the magnons 
is then $\lambda_M=t/(2J)=1.67$ for $J/t=0.3$ while the coupling 
to phonons is only $\lambda=1.2$ or 0.75 according to the estimates
in Sec.~\ref{sec:epp}. It is then interesting to ask why not spin    
fluctuations alone can drive polaron formation. 

Mishchenko and Nagaosa \cite{Mishchenko} and as well as Ciuchi {\it et al.}
\cite{Ciuchi97,Cappelluti} pointed out that in a diagrammatic
description, polaron formation requires the inclusion of diagrams 
where lines describing phonon Green's functions cross. Liu and 
Manousakis \cite{ZL92} showed that due to symmetry reasons, whole 
classes of such diagrams are identically zero for coupling to magnons. 
This should be an essential reason why magnons alone cannot lead to 
polarons in the sense that the quasiparticle weight goes exponentially 
to zero as is the case for coupling to phonons. Alternatively, one can notice 
that polaron formation for Holstein phonons involves the excitation
of many phonons on the same site as the hole. The magnons are
due to flipping spins with $s=1/2$, and locally it is only possible
to flip such a spin once. This is in contrast to local phonons,
which are true bosons and can be excited infinitely many times.  

\subsection{Kinks}\label{eq:ek}

Much of the interest in the EPI in the context of cuprates was triggered
as Lanzara {\it et al.} \cite{Lanzara2001} emphasized the presence of 
kinks in the dispersion of the photoemission spectrum (see Fig.~\ref{fig:el1}).
There have been a large number of theoretical studies of this effect
\cite{Norman2000,Zeyher2001,Manske2001,Norman2002,Manske2003,Norman2004,Sandvik,Devereaux2004,Manske2004}. An extensive theoretical study was performed by Sandvik 
{\it et al.} \cite{Sandvik}, who solved the Eliashberg equations approximately,
considering coupling to Holstein, breathing and buckling phonons as 
well as to the resonance peak seen in neutron scattering. The ${\bf q}$
dependence of the coupling for these modes is assumed to be rather 
different. The calculated spectra, however, did not show large qualitative
differences. It was concluded that from the ${\bf q}$ dependence alone, 
it might be hard to determine which mode causes the main coupling \cite{Sandvik}.

Devereaux {\it et al.} \cite{Devereaux2004} performed similar calculations,
focusing on the coupling to breathing (at $\omega_{\rm Br}=70$ meV) and 
B$_{1g}$ buckling (at $\omega_{B_{1g}}=36$ meV) phonons and comparing with
experimental results for Bi$_2$Sr$_2$Ca$_{0.92}$Y$_{0.08}$Cu$_2$O$_{8+\delta}$. 
They argued that the B$_{1g}$ phonon in particular couples to the antinodal 
point while the breathing phonon couples mainly to the nodal point. In the 
superconducting state, they found a structure in the antinodal direction at about 
$\omega_{B_{1g}}+\Delta=71$ meV, where $\Delta=35$ meV is the gap, while in 
the nodal direction they found a structure at about $\omega_{\rm Br}=70$ meV. 
The effects of the $c$-axis O phonons on the kink has been studied by 
Meevasana {\it et al.} \cite{Meevasana1,Meevasana2}, as discussed in 
Sec.~\ref{sec:ica}. 

Norman {\it et al.} \cite{Norman1997} noticed that the peak-dip-hump feature 
seen at the antinodal point below $T_c$ could be explained by a coupling 
to the resonance peak, since this peak also only appears below $T_c$, has 
the right energy and would couple particularly strongly to the antinodal 
point. It was concluded that there may be a strong coupling to this resonance. 
This led to a substantial amount of work 
\cite{Norman2000,Manske2001,Norman2002,Manske2003,Norman2004,Manske2004}
describing a kink in terms of the coupling to spin fluctuations and to the
resonance peak. 

The kink is particularly pronounced in the anti-nodal direction in multilayer 
systems below $T_\mathrm{c}$. The resonance peak is observed under these 
conditions, and it may therefore contribute to the kink. For symmetry reasons, 
the B$_{1g}$ buckling phonon has a substantial coupling for multilayer 
systems \cite{Devereaux2004}. The pile up of density of states around the 
superconducting gap should enhance the contribution of the B$_{1g}$ phonon to 
the kink for $T<T_\mathrm{c}$ \cite{Devereaux2004}. This makes it hard to 
determine the relative importance of the resonance peak and the B$_{1g}$ 
phonon for the kink in the anti-nodal direction in multilayer systems. 

Eschrig and Norman \cite{Norman2002} observed that since the resonance peak 
is seen for bilayer systems in the odd channel it couples bonding (with 
respect to the two layers) to antibonding states. Since there is a large 
density of antibonding states close to the Fermi energy, the 
resonance peak should in particular influence the bonding state. They 
concluded \cite{Norman2002} that the spectra of Bi2212 close to the 
antinodal point, interpreted in terms of antibonding and bonding states 
with a peak-dip-hump structures \cite{Feng2001}, could be described if a 
sharp mode was introduced in the odd channel but not in the even channel. 
An interesting question is what happens if there is a sharp mode in both 
channels, as expected for phonons in the CuO$_2$ plane.  Borisenko {\it et 
al.} \cite{Borisenko2006} applied the same idea to the spectra close to the 
nodal point. They found that the width of a bonding state is larger than for 
an antibonding state of the same energy (but different ${\bf k}$), and 
concluded that this favors coupling in the odd channel. If, however, the 
broadening at zero binding energy (perhaps due to impurity scattering and 
other elastic effects) is subtracted, the conclusion is actually reversed. 

\subsection{Isotope effect}\label{sec:ei}

There have been a number of theoretical studies of the isotope
effect \cite{Andergassen,Paci,Seibold,Fratini,Mishchenko2006,Paci2006}.
Andergassen {\it et al.} \cite{Andergassen} studied the phase
diagram related to a quantum critical point (QCP) due to incommensurate
charge ordering. Going beyond a mean-field theory by including 
fluctuations, they found that the QCP can be shifted by the 
isotope effect, leading to a general shift of the phase diagram. 
This shows up as a strong isotope effect in various properties
\cite{Andergassen}. Seibold and Grilli \cite{Seibold} argued that
the correlation length for charge order fluctuations might have
an isotope effect and found that this can lead to isotope effects 
in the PES spectra. 
Paci {\it et al.} \cite{Paci} studied the Holstein model and 
demonstrated a strong isotope effect in, e.g., the effective mass 
when the EPI was strong enough to put the system close to (small)
polaron formation. In a study of the Holstein-Hubbard model,
Paci {\it et al.} \cite{Paci2006} found that the competion between 
the Coulomb and electron-phonon interactions strongly influences
the isotope effect.  Mishchenko and Nagaosa \cite{Mishchenko2006}
studied the undoped Holstein-$t$-$J$ model using a diagrammatic 
Monte-Carlo method. They found that the isotope effect can be large
under certain circumstances. 

\section{Superconductivity}\label{sec:es}

Honerkamp {\it et al.} \cite{Honerkamp1,Honerkamp2} have studied
the effects of phonons on the superconductivity transition using 
a weak-coupling functional renormalization group method. They
studied Holstein, breathing, A$_{1g}$ and B$_{1g}$ out-of-plane 
oxygen buckling modes. They found that the Holstein and A$_{1g}$
phonons are unfavorable for $d$-wave superconductivity. These 
phonons add an on-site interaction, which reduces the Coulomb 
repulsion.  In the weak-coupling limit studied by Honerkamp {\it et al.} 
\cite{Honerkamp1,Honerkamp2}, this was found to be unfavourable for
d-wave superconductivity. Of the phonons studied, only the B$_{1g}$ 
phonon, which gives no on-site attraction for ${\bf q}=0$, was found 
to be favorable for $d$-wave superconductivity.

\begin{figure}[bt]
\centerline{
{\rotatebox{0}{\resizebox{10.0cm}{!}{\includegraphics {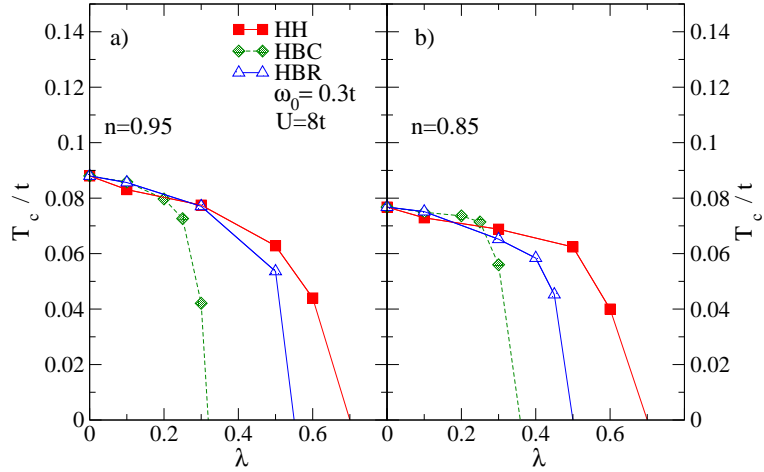}}}}}
\caption[]{\label{fig:es1}$T_\mathrm{c}$ versus $\lambda$ for for a Hubbard
model with Holstein (HH), buckling (HBC) or breathing (HBR) phonons at 
(a) 5$\%$ and (b) 15$\%$ doping for $\omega_\mathrm{ph}=0.3 t$ and $U=8t$ 
(after Macridin {\it et al.} \cite{Macridin2}). 
 
}
\end{figure}

Macridin {\it et al.} \cite{Macridin1,Macridin2} performed DCA calculations 
for Hubbard model with coupling to Holstein, breathing and out-of-plane
oxygen buckling phonons for a $2 \times 2$ cluster. The calculations 
neglected the EPI on the hopping integrals and thereby the ${\bf k}$
dependence of the coupling. The results for the transition temperature 
$T_\mathrm{c}$ are shown in Fig.~\ref{fig:es1}. The figure shows how 
$T_\mathrm{c}$ drops 
with $\lambda$ for all the phonon couplings considered. It was actually 
found that all phonons considered enhance the pairing, and it was 
speculated that this is due to an enhancement of the AF susceptibility 
\cite{Macridin2}. The EPI, however, reduces the quasiparticle strength $Z$, 
and it was concluded that this is more important than the increase of
the pairing, leading to a reduction of $T_\mathrm{c}$ \cite{Macridin2}. It would 
be interesting to also study the effects of coupling to hopping integrals for 
the buckling modes and the effects for a larger $U$.

\section{Summary}\label{sec:sum}

We have reviewed evidence that the electron-phonon interaction (EPI)
can substantially influence various properties of the high-$T_c$
cuprates. Some of the evidence is indirect and it can then be hard to
determine which phonon modes are involved or to distinguish between 
phonons and other bosonlike excitations, such as the resonance peak.
It is therefore of particular interest that 
inelastic neutron scattering shows that certain phonon modes are appreciably
broadened and shifted when the system is doped. This is strong evidence 
that these modes have a substantial EPI. In particular, such effects have 
been found for apical oxygen phonons and the (half-)breathing oxygen
bond-stretching phonons. Theoretical many-body calculations indeed find 
substantial couplings for these phonons. For multilayer systems similar 
effects are also seen for other phonons, in particular, the B$_{1g}$ 
phonon. We have discussed theoretical treatments of phonon softening and 
broadening and illustrated that the experimental behavior can be described 
theoretically.  

Much of the interest in the EPI was triggered by the observation of 
kinks in the dispersion seen in photoemission, which can be interpreted
in terms of coupling to phonons.  It remains controversial, however,  
how much phonons and the resonance peak or other spin excitations 
contribute to these kinks. Theoretical treatments of the kink have been shown. 

There is strong evidence in favor of (small) polaron formation in undoped
cuprates.  While this polaron formation is helped by the antiferromagnetic
correlations, we have argued that antiferromagnetic fluctuations alone 
could not lead to small polarons. This suggests a substantial EPI 
for the undoped cuprates. The EPI for LaCuO$_4$ has been calculated,
obtaining substantial coupling to apical and breathing phonons, 
and it was shown that the experimental line shape of the photoemission spectra 
can be understood rather well. As the system is doped, the polarons disappear. 
Photoemission spectra, however, still show substantial weight in the energy 
range where the polaron related phonon sideband was observed, suggesting that 
the EPI is substantial also in the doped systems.  

An important feature of the cuprates is the great importance of the 
Coulomb repulsion. We have therefore treated the interplay between Coulomb 
and electron-phonon interactions extensively. Using sum rules, we showed
that for weakly doped cuprates, the Coulomb repulsion strongly suppresses 
the phonon self-energy, while there is no corresponding strong suppression 
of the electron self-energy. For polaron formation as seen in photoemission 
(described by the electron self-energy), antiferromagnetic correlations, 
resulting from the Coulomb repulsion,  greatly help the EPI. Nevertheless, 
due to other effects of the Coulomb interaction, there is a moderate 
suppression of polaron formation, at least in the antiferromagnetic DMFT.

The EPI is usually discussed for some model with EPI, assuming that the
electron-phonon coupling constants are fixed. We find, however, that
when deriving, for instance,  Holstein-$t$-$J$ or Holstein-Hubbard models,
the coupling constants may be substantially enhanced by many-body effects.
This is, for instance, the case for the (half-)breathing phonons.

\end{document}